\documentclass{jfm}
\usepackage{natbib}
\usepackage[dvips]{graphicx}
\usepackage{psfrag}
\usepackage{amsmath}

\ifCUPmtlplainloaded \else
  \IfFileExists{amsbsy.sty}
    {\typeout{^^JFound the 'amsbsy' package on the system, using it.^^J}%
     \usepackage{amsbsy}}
    {\providecommand\boldsymbol[1]{\mbox{\boldmath $##1$}}}
\fi

\renewcommand{\vec}[1]{\boldsymbol{#1}}

\usepackage{color}
\newcommand{\difft}[1]{\partial_t #1}
\newcommand{\diffz}[1]{\partial_z #1}
\newcommand{\diffZ}[1]{\f{\partial  #1}{\partial Z}}
\newcommand{\diffzz}[1]{\partial^2_z #1}
\newcommand{\diffZZ}[1]{\f{\partial^2 #1}{\partial Z^2}}
\newcommand{\diffzp}[1]{\partial'_z #1}
\newcommand{\diffzzp}[1]{\partial'^2_z #1}
\newcommand{\gradx}[2]{\partial_{#2} #1}
\newcommand{\gradxp}[2]{\partial'_{#2} #1}
\newcommand{\gradX}[2]{\f{\partial #1}{\partial X_{#2}}}
\newcommand{\gradr}[2]{\f{\partial #1}{\partial r_{#2}}}
\newcommand{\lapx}[2]{\partial^2_{#2} #1}
\newcommand{\lapxp}[2]{\partial'^2_{#2} #1}
\newcommand{\lapr}[2]{\f{\partial^2 #1}{\partial r_{#2}^2}}

\newcommand{\moy}[1]{\left<#1\right>}
\newcommand{\f}[2]{\frac{#1}{#2}}

\newcommand{\pr}{Pr}

\newcommand{\ray}{Ra}
\newcommand{\rey}{Re}
\newcommand{\nuss}{Nu}
\newcommand{\aspra}{A}

\newlength{\hauteur}
 \newcommand{\point}{\protect\rule[\hauteur]{0.7pt}{0.6pt}}
\newcommand{\spa}{\hspace{1.2mm}}
\newcommand{\bigdash}{\protect\rule[\hauteur]{2.4mm}{0.5pt}}
\newcommand{\smalldash}{\protect\rule[\hauteur]{1.mm}{0.5pt}}
\newcommand{\fullline}{\protect\rule[\hauteur]{1cm}{0.5pt}}
\newcommand{\longdashline}{\bigdash\spa\bigdash\spa\bigdash}
\newcommand{\dashline}{\smalldash\spa\smalldash\spa\smalldash\spa\smalldash\spa\smalldash}
\newcommand{\dotline}{\point\spa\point\spa\point\spa\point\spa\point\spa\point\spa\point\spa\point}
\newcommand{\dashdotline}{\smalldash\spa\point\spa\smalldash\spa\point\spa\smalldash\spa\point} 
\newcommand{\dashtripledotline}{\bigdash\spa\point\spa\point\spa\point\spa\bigdash}

\title[Anisotropy, inhomogeneity and inertial range scalings in
  turbulent convection]{Anisotropy, inhomogeneity and inertial range scalings in
  turbulent convection}

\author[F. Rincon]{F\ls R\ls A\ls N\ls \c{C}\ls O\ls I\ls S\ns R\ls I\ls N\ls C\ls O\ls N}
\affiliation{Observatoire Midi-Pyr\'en\'ees, UMR 5572, Universit\'e Paul Sabatier et CNRS,\\ F-31400
 Toulouse, France}

\begin{document}
\maketitle
\setcounter{tocdepth}{3}
\begin{abstract}
This paper provides a detailed study of turbulent statistics and scale-by-scale
budgets in turbulent Rayleigh-B\'enard convection. It aims at testing
the applicability of \cite{k41} and \cite{bolgiano59}
theories in the case of turbulent convection and at improving the understanding
of the underlying inertial range scalings, for which a general agreement is still lacking.
Particular emphasis is laid on anisotropic and inhomogeneous
effects, which are often observed in turbulent convection between two
differentially heated plates. For this purpose, the SO(3) decomposition of structure
functions \citep*{arad99a} and a method of description of
inhomogeneities  proposed by \cite{danaila01} are used
to derive inhomogeneous and anisotropic generalizations of Kolmogorov and Yaglom
equations applying to Rayleigh-B\'enard convection, which
can be extended easily to other types of anisotropic and/or inhomogeneous flows. 
The various contributions to these equations are computed in and off
the central plane of a convection cell using data produced by a
Direct Numerical Simulation of turbulent Boussinesq convection at
$\ray=10^6$ and $\pr=1$ with aspect ratio $A=5$. 
The analysis of the isotropic part of the
Kolmogorov equation demonstrates that the shape of the third-order velocity
structure function is significantly
 influenced by buoyancy forcing and large-scale inhomogeneities, 
while the isotropic part of the mixed third-order structure function
 $\moy{(\Delta\theta)^2\Delta\vec{u}}$ appearing in Yaglom equation exhibits
 a clear scaling exponent 1 in a small range of scales. 
The magnitudes of the various low $\ell$ degree anisotropic components
of the equations are also estimated and  are  shown to be comparable
to their isotropic counterparts at moderate to large scales. 
The analysis of anisotropies notably reveals that computing
reduced structure functions (structure functions computed at fixed depth
for correlation vectors $\vec{r}$ lying in specific planes only) in
order to reveal scaling exponents predicted by isotropic theories is
misleading in the case of fully three-dimensional turbulence in the bulk
of a convection cell, since such quantities involve linear combinations
of different $\ell$ components which are not negligible
in the flow. This observation also indicates
that using single points measurements together with the Taylor hypothesis in the
particular direction of a mean flow to test the predictions of
asymptotic dimensional isotropic theories of turbulence
or to calculate intermittency corrections to these theories may lead
to significant biases for mildly anisotropic three-dimensional flows.
A qualitative analysis is finally used to show
that the influence of buoyancy forcing at scales smaller than the
Bolgiano scale is likely to remain important up to
$\ray=10^9$, thus preventing Kolmogorov scalings from showing
up in convective flows at lower Rayleigh numbers. 
\end{abstract}

\section{Introduction}
The quest for inertial range scaling laws in 
turbulent convection has been very active in recent years
(\textit{e.~g.}
\cite{chilla93,benzi94,calzavarini02,verzicco03,ching04}). 
Their determination is expected to give some important
insight into the thermal and mechanical processes at work in the
flow. To this end, turbulent thermal convection is investigated 
in convection cells heated from below using
laboratory and numerical experiments, within the
framework of the Boussinesq approximation. Such a flow exhibits two
essential properties: it is both strongly anisotropic and
inhomogeneous. Anisotropy comes from gravity, while inhomogeneity
results from the presence of top and bottom horizontal boundaries in convection cells.
As a consequence, inertial range scalings of convective turbulence, when they can
ever be observed, depend  strongly on the vertical coordinate.  

Accordingly, asymptotic theories of turbulence constructed under the
assumptions of homogeneity and isotropy may be partially irrelevant to
understand the observed properties of turbulent convection. 
Neglecting intermittency effects, the classical picture regarding
scaling laws in this flow is that Bolgiano-Obukhov turbulence (\cite{bolgiano59,obukhov59},
hereafter BO59 theory) should be present on correlation lengths  $r$
larger than the so-called Bolgiano length. A dimensional estimate of
this length \citep{chilla93} is given by
\begin{equation}
\label{bolglength}
  L_B=\frac{\nuss^{1/2}d}{(\ray\pr)^{1/4}}~,
\end{equation}
where $d$ is the depth of the convective layer, $\nuss$ is the Nusselt number,
$\ray$ is the Rayleigh number and $\pr$ is the Prandtl number. Instead,
homogeneous and isotropic Kolmogorov turbulence (\cite{k41}, hereafter
K41 theory) should be observed for $r<L_B$.
In the Bolgiano-Obukhov subrange, a dominant balance between buoyancy forcing (for an
unstably stratified layer) and the third-order structure
function occurs, which modifies the turbulent energy cascade
substantially in comparison to K41: while longitudinal velocity
increment scalings $\moy{\Delta u_r(r)}\sim r^{1/3}$ are predicted
asymptotically in the inertial range for K41,
they should follow $\moy{\Delta u_r(r)}\sim r^{3/5}$ in the BO59
regime. As far as temperature increments are concerned,
$\moy{\Delta \theta(r)}\sim r^{1/3}$ results from K41 (passive
scalar scalings), while $\moy{\Delta \theta(r)}\sim r^{1/5}$ is expected
from dimensional arguments in BO59 theory.
These different regimes should in principle be detected
through the measurements of structure functions or spectra when
turbulence is sufficiently developed. However, clear evidences of K41
or BO59 scalings are still lacking in both numerical and experimental
convection, even at very high Rayleigh numbers.
Mixed scalings, which are not compatible with dimensional analysis,
have sometimes been detected (\textit{e. g.} K41 scalings
for the velocity and BO59 scalings for temperature in the same
wave number range, see for instance
\cite{verzicco03}).
BO59 scalings have been reported in the bulk of experimental convection
cells \citep{benzi94,ching04}, and in a numerical experiment 
\citep*{calzavarini02}, close to the horizontal walls of the
convection cell. Direct scaling laws could not be observed for reduced structure
functions (structure functions computed at fixed depth for
correlation vectors $\vec{r}$ lying in horizontal planes only) in the
latter study, so that the derivation of scaling exponents by \cite{calzavarini02} relies on
Extended Self Similarity (ESS, see \cite{benzi93}).  Results obtained for
various anisotropic flows, such as the channel flow \citep{arad99b},
demonstrate that it is sometimes possible to observe scaling laws
by plotting structure functions directly, even at modest Reynolds
numbers, provided that the complete computation and SO(3) decomposition of structure
functions are performed: in the work of \cite{arad99b}, the difference
between reduced structure functions
and the isotropic (angular mean) component of the complete structure
functions is striking.  This spherical harmonics decomposition
of structure functions has also been computed by 
\cite{biferale03} for homogeneous Rayleigh-B\'enard
convection (HRB) in a regime where $L_B$ is comparable to the box size,
thus preventing any possibility of BO59 scalings. The anisotropic scaling
exponents reported for HRB are anomalous, which means that they do not
match dimensional analysis predictions either. The understanding
of inertial range scalings in convective turbulence and their very existence
even at high Rayleigh numbers therefore remains a puzzling problem that
still deserves important efforts. 

This paper aims at presenting a complete description of turbulent statistics
in Boussinesq convection  at $\ray=10^6$ and $\pr=1$ in order to test explicitly
the assumptions of both K41 and BO59 theories in such a flow and to try
to  clarify some of the previously mentioned  problems regarding the occurrence of
inertial range scaling laws in convective turbulence. 
The study shows that inhomogeneity, anisotropy
and the presence of buoyancy forcing at all scales all contribute
to the absence of scaling laws, and points out that very
high Rayleigh numbers should be reached in order to be able to observe the
definite signature of spectral scalings in turbulent convection. 
The presentation of the results is as follows. The SO(3) decomposition
of statistical averages is first used in \S\ref{theorie} to derive
generalized Kolmogorov and Yaglom equations including
inhomogeneous and anisotropic terms, which describe
respectively the production, transport and dissipation of velocity and
temperature fluctuations. A detailed numerical study of these
equations  is then proposed using Direct Numerical Simulation (DNS) data. 
The DNS and data processing algorithms are first described
in \S\ref{numdetails}. The various contributions to the
equations are then presented in \S\ref{results} at the
center of the convection cell and off the central plane, with particular
emphasis laid on the anisotropic and inhomogeneous effects inferred from the analysis.
The consequences and perspectives offered by these results are finally
discussed in \S\ref{discuss}.

\section{\label{theorie}Theoretical considerations}
Under the assumptions of homogeneity and isotropy, \cite{k41} derived
his famous equation for the third-order structure function of the
turbulent velocity field
\begin{equation}
\label{eqkolmo}
  \moy{(\Delta u_r)^3}=-\f{4}{5}\moy{\varepsilon} r+6\nu\gradr{}{}\moy{(\Delta u_r)^2}~,
\end{equation}
which leads to the well-known K41 scaling law 
in the inertial range of turbulence. To obtain it, one also has to assume 
that the forcing mechanism of turbulence is important on large
scales only. There are however numerous flows, such as
Rayleigh-B\'enard convection, where some of these
assumptions possibly break down. 
A detailed study of scale-by-scale energy budgets in
turbulent convection is therefore required to test 
explicitly the accuracy of these assumptions. 
Indeed, the velocity budget will differ significantly from
equation~(\ref{eqkolmo}) if one
or several of these hypothesis are violated. The objective of this
section is to provide a derivation of 
generalized forms of Kolmogorov and Yaglom equations including
anisotropic and inhomogeneous effects, in order to be able to perform
comparisons with equation~(\ref{eqkolmo}) and with the homogeneous
and isotropic version of Yaglom equation, for a given convective flow.
A procedure inspired by \cite{monin75}, \cite{lindborg96},
\cite{hill97}, \cite{antonia97} and \cite{hill2002}, is adopted here. The method has
already been applied to channel flow turbulence \citep{danaila01}, heated
decaying turbulence \citep{danaila99} and shear turbulence
\citep{casciola03}, which, as Rayleigh-B\'enard convection, are
anisotropic (and sometimes inhomogeneous) flows. Inhomogeneity is
investigated in a similar fashion as in the paper by \cite{danaila01}.
The procedure is supplemented by a spherical harmonics decomposition
of structure functions \citep{arad99a} which serves to quantify
anisotropic effects.
It should be noted that a study of structure function scalings in
turbulent convection is available  \citep{yakhot92} and that it has
been used by several authors (\textit{e. g.}
\cite*{benzi98}) to predict scaling behaviour in numerical
experiments.  However, this work assumes isotropy and homogeneity,
which, as will be shown later in the paper, are not appropriate for
this flow. 

\subsection{Derivation of  the inhomogeneous Kolmogorov and Yaglom equations}
Summation over repeated indices is assumed throughout the derivation. 
A Boussinesq fluid of constant density $\rho$, kinematic viscosity
$\nu$ and thermal diffusivity $\kappa$ is used. Gravity is denoted by
$\vec{g}=-g\vec{e}_z$, where $z$ is the vertical coordinate.
The Boussinesq equations \citep{chandra61} for the velocity field
$\vec{u}$, pressure field $p$ and temperature field $T$ (separated
into a mean part $\moy{T}$ and fluctuations $\theta$)  read 
\begin{equation}
  \label{eqbouss1}
  \difft{u_i}+u_k\gradx{u_i}{k}=-\f{1}{\rho}\gradx{p}{i}+\alpha
  g\,\theta\,\delta_{iz}+\nu\lapx{u_i}{k}~,
\end{equation}
\begin{equation}
  \label{eqbouss2}
  \difft{\moy{T}}+\difft{\theta}+u_k\gradx{\theta}{k}+u_z\diffz{\moy{T}}=\kappa\lapx{\theta}{k}+\kappa\diffzz{\moy{T}}~,
\end{equation}
\begin{equation}
  \label{eqdiv}
  \gradx{u_k}{k}=0
\end{equation}
at position $\vec{x}$ and
\begin{equation}
  \label{eqbouss1p}
  \difft{u_i'}+u_k'\gradxp{u_i'}{k}=-\f{1}{\rho}\gradxp{p'}{i}+\alpha
  g\,\theta'\,\delta_{iz}+\nu\lapxp{u_i'}{k}~,
\end{equation}
\begin{equation}
  \label{eqbouss2p}
\difft{\moy{T'}} + \difft{\theta'}+u_k'\gradxp{\theta'}{k}+u_z'\diffzp{\moy{T'}}=\kappa\lapxp{\theta'}{k}+\kappa\diffzzp{\moy{T'}}~,
\end{equation}
\begin{equation}
  \label{eqdivp}
  \gradxp{u_k'}{k}=0
\end{equation}
at position $\vec{x}'$ (primes denote quantities evaluated at
$\vec{x}'$). As is usual in this kind of analysis, $\vec{x}$ and $\vec{x}'$ are independent
variables so that $\partial_k u_i'=\partial_k' u_i=0$.
Subtraction of equations (\ref{eqbouss1}) from (\ref{eqbouss1p}) and of
(\ref{eqbouss2}) from (\ref{eqbouss2p}) thus leads to
\begin{align}
  \label{eqboussdelta}
\qquad \difft{\Delta u_i}+\Delta
  u_k\gradxp{\Delta u_i}{k}+u_k\left(\gradx{}{k}+\gradxp{}{k}\right)\Delta
  u_i =  \qquad\qquad\qquad\qquad\qquad \notag\\  -\f{1}{\rho}\left(\gradx{}{i}+\gradxp{}{i}\right)\Delta
  p + \alpha  g\,\Delta\theta\,\delta_{iz} + \nu \left(\lapx{}{k}+\lapxp{}{k}\right)\Delta u_i~,
\end{align}
\begin{align}
  \label{eqbouss2delta}
 \difft{\Delta \moy{T}}+\difft{\Delta \theta}+\Delta u_k\gradxp{\Delta \theta
    }{k}+u_k\left(\gradx{}{k}+\gradxp{}{k}\right)\Delta
  \theta\notag  +u_z'\diffzp{\moy{T'}}-u_z\diffz{\moy{T}}=\qquad\qquad\notag\\ \kappa\left(\lapx{}{k}+\lapxp{}{k}\right)\Delta \theta+
\kappa\left(\lapxp{\moy{T'}}{z}-\lapx{\moy{T}}{z}\right)~,
\end{align}

\noindent where  $\Delta f= f'-f$ for any variable $f$.
Multiplying  equations (\ref{eqboussdelta}) and
(\ref{eqbouss2delta}) by $2\Delta u_i$
and $2\Delta \theta$ respectively, using incompressibility 
and averaging gives
\begin{align}
  \label{eqboussdeltaav}
  \difft{}\moy{(\Delta u_i)^2}+\moy{\gradxp{}{k}\left(\Delta
  u_k(\Delta u_i)^2\right)}+\moy{\left(\gradx{}{k}+\gradxp{}{k}\right)\left(u_k(\Delta
  u_i)^2\right)}=\qquad\qquad\qquad\qquad\qquad\notag  \\  -\f{2}{\rho}\moy{\left(\gradx{}{i}+\gradxp{}{i}\right)\left(\Delta
  u_i\Delta p\right)}+ 2\alpha  g\,\moy{\Delta\theta\Delta
u_z}+\nu \moy{\left(\lapx{}{k}+\lapxp{}{k}\right)(\Delta
u_i)^2} - 2\left(\moy{\varepsilon} + \moy{\varepsilon'}\right)~,
\end{align}
\begin{align}
  \label{eqbouss2deltaav}
\difft{}\moy{(\Delta \theta)^2}+\moy{\gradxp{}{k}\left(\Delta
     u_k(\Delta
     \theta)^2\right)}+\moy{\left(\gradx{}{k}+\gradxp{}{k}\right)\left(u_k(\Delta
  \theta)^2\right)}\qquad\qquad\qquad\notag\phantom{\diffZ{}} \\ +2\moy{u_z'\diffzp{\moy{T'}}\Delta \theta}-
2\moy{u_z\diffz{\moy{T}}\Delta\theta}=\qquad\qquad\notag\phantom{\diffZ{}}\\
\kappa\moy{\left(\lapx{}{k}+\lapxp{}{k}\right)(\Delta \theta)^2}
-2\left(\moy{N}+\moy{N'}\right)~,
\end{align}
where $\moy{\varepsilon}=\nu\moy{(\partial_iu_j)^2}$ and $\moy{N}=\kappa\moy{(\partial_i\theta)^2}$.
Following \cite{hill2002}, the analysis can be pushed further by using new variables
$\vec{X}=(\vec{x}+\vec{x}')/2$ and $\vec{r}=\vec{x}'-\vec{x}$.
Differential operators transform according to
\begin{equation}
 \begin{array}{ccrcl}
  \displaystyle{\gradx{}{k}} & = & \displaystyle{-\gradr{}{k}} &+ &
  \displaystyle{\f{1}{2}\gradX{}{k}}~, \\ \\
  \displaystyle{\gradxp{}{k}} & = & \displaystyle{\gradr{}{k}} & + & \displaystyle{\f{1}{2}\gradX{}{k}}~.
\end{array}
\end{equation}
As the Rayleigh-B\'enard flow is homogeneous in the horizontal
directions, the $\vec{X}$ dependence  is restricted to $Z$, 
so that only $\partial_Z$ derivatives have
to be considered. It is important to outline that $\partial_Z$ is not
the same operator as $\diffz{}$, since  $\vec{r}$ is to be kept
constant when the former is applied, while $z'$ must be kept constant
in the latter case. Assuming statistical stationarity,
equations~(\ref{eqboussdeltaav})-(\ref{eqbouss2deltaav}) become
 \begin{align}
   \label{eqboussdeltaavXr}
   \gradr{}{k}\moy{\Delta u_k(\Delta u_i)^2}+\f{1}{2}\diffZ{}\moy{\Delta u_z(\Delta u_i)^2}+\diffZ{}\moy{u_z(\Delta
   u_i)^2}=\qquad\qquad\qquad\notag\\  -2\left(\moy{\varepsilon}+\moy{\varepsilon'}\right)
-\f{2}{\rho}\diffZ{}\moy{\Delta u_z\Delta p}+  2\alpha
g\,\moy{\Delta u_z\Delta\theta}\qquad\notag\\+2\nu\lapr{}{k}\moy{(\Delta u_i)^2}+\f{\nu}{2}\diffZZ{}\moy{(\Delta u_i)^2}~,
 \end{align}
 \begin{align}
   \label{eqbouss2deltaavXr}
\gradr{}{k}\moy{\Delta u_k(\Delta \theta)^2} +\f{1}{2}\diffZ{}\moy{\Delta u_z (\Delta \theta)^2}+\diffZ{}\moy{u_z(\Delta
         \theta)^2} \qquad\qquad\qquad\qquad\qquad\qquad\notag\\+2\moy{u_z'\diffzp{\moy{T'}}\Delta \theta}
-2\moy{u_z\diffz{\moy{T}\Delta\theta}}=\qquad\qquad\qquad\qquad\notag\phantom{\diffZ{}}\\
       -2\left(\moy{N}+\moy{N'}\right)
+2\kappa\gradr{}{k}\moy{(\Delta \theta)^2}+\f{\kappa}{2}\diffZZ{}\moy{(\Delta\theta)^2}~,
 \end{align}
where all quantities now depend on $(Z,\vec{r})$.
\subsection{Dealing with anisotropy}
To analyse the effects of anisotropy, 
a decomposition of the $\vec{r}$-dependence of statistical averages on
spherical harmonics must be performed
\citep{arad99a}. For scalar averages, denoted here by $\moy{F}$, such as $\moy{(\Delta
  u_i)^2}$, this decomposition reads
\begin{equation*}
  \label{eqylm}
  \moy{F}(Z,\vec{r})=\sum_{\ell=0}^{+\infty}\sum_{m=-\ell}^{+\ell}\moy{F}^{\ell}_m(Z,r)Y^m_{\ell}(\theta,\varphi)~,
\end{equation*}
where $Y^m_\ell(\theta,\varphi)$ denotes the spherical harmonic of sectorial order
$\ell$ and azimuthal order $m$  (here, $\theta$ denotes the angle between
the vertical direction $z$ and $\vec{r}$ and should not be confused
with a temperature perturbation).
One also has to deal with the vectorial averages $\moy{\vec{U}}\equiv\moy{(\Delta
  u_i)^2\Delta\vec{u}}$ and
$\moy{\vec{\Theta}}\equiv\moy{(\Delta \theta)^2\Delta\vec{u}}$
  appearing in equations~(\ref{eqboussdeltaavXr})-(\ref{eqbouss2deltaavXr}). The
spherical harmonics representations of these vectors are
\begin{align*}
  \moy{\vec{U}}(Z,\vec{r})= \sum_{\ell=0}^{+\infty}\sum_{m=-\ell}^{+\ell}
                            \moy{U_R}^{\ell}_m(Z,r)\vec{R}^m_{\ell}(\theta,\varphi)\notag\qquad\qquad\qquad\qquad\qquad\qquad\\
                            +\moy{U_S}^{\ell}_m(Z,r)\vec{S}^m_{\ell}(\theta,\varphi)+
                            \moy{U_T}^{\ell}_m(Z,r)\vec{T}^m_{\ell}(\theta,\varphi)~,
\end{align*}
\begin{align*}
  \moy{\vec{\Theta}}(Z,\vec{r})= \sum_{\ell=0}^{+\infty}\sum_{m=-\ell}^{+\ell}
                            \moy{\Theta_R}^{\ell}_m(Z,r)\vec{R}^m_{\ell}(\theta,\varphi)\notag\qquad\qquad\qquad\qquad\qquad\qquad\\
                            +\moy{\Theta_S}^{\ell}_m(Z,r)\vec{S}^m_{\ell}(\theta,\varphi)+
                            \moy{\Theta_T}^{\ell}_m(Z,r)\vec{T}^m_{\ell}(\theta,\varphi)~,
\end{align*}
where $\vec{R}^m_{\ell}=Y^m_{\ell}\vec{e}_{r}$, $\vec{S}^m_{\ell}=\vec{\nabla}Y^m_{\ell}$,
$\vec{T}^m_{\ell}=\vec{\nabla}\times\vec{R}^m_{\ell}$, and $\vec{\nabla}$ is
defined at $r=1$. $\moy{U_R}^\ell_m$ and $\moy{U_S}^\ell_m$
are the components of a vector projected directly on a spherical harmonics
vectorial basis. They are in that sense different from a quantity such
as $\moy{\Delta u_z\Delta \theta}^\ell_m$, which results from a projection on
  $\vec{e}_z$ followed by a projection on spherical harmonics. However,
  note that  $\moy{U_R}^\ell_m$ could be noted $\moy{(\Delta
  u_i)^2\Delta u_r}^\ell_m$ as well, since $\vec{R}^m_{\ell}$ is along $\vec{e}_r$.

As the spherical harmonics form an orthogonal basis,
equations for different $\ell$ and $m$ can be considered separately.
Also, a dependence of statistical averages on $Z$ and $r$ only will be
assumed from now on and the $(Z,r)$ notation will be
omitted subsequently. Using this formalism, the divergence operator
acting for instance on $\moy{\vec{U}}$ reads 
  \begin{equation*}
    \left(\gradr{\moy{U_k}}{k}\right)^{\ell}_m=\f{1}{r^2}\gradr{}{}\left(r^2\moy{U_R}^{\ell}_{m}\right)
                         -\f{\ell(\ell+1)}{r}\moy{U_S}^{\ell}_m~,
  \end{equation*}
while the laplacian of a scalar average is given by
\begin{equation*}
  \left(\lapr{\moy{F}}{k}\right)^{\ell}_m=\f{1}{r^2}\gradr{}{}\left(r^2\displaystyle{\gradr{}{}\moy{F}^{\ell}_m}\right)-\f{\ell(\ell+1)}{r^2}\moy{F}^{\ell}_m~.
\end{equation*}
Projecting equations
(\ref{eqboussdeltaavXr})-(\ref{eqbouss2deltaavXr}) on $Y^m_{\ell}$
for all $\ell$ and $m$ and applying the operator $\displaystyle{1/r^2\int_0^ry^2
  \bullet \mbox{d}y}$ to the result, the following hierarchy of
equations is readily obtained:
\begin{equation}
\label{eqVfinal}
\begin{array}{rl}
\displaystyle{\moy{U_R}^{\ell}_m-\f{\ell(\ell+1)}{r^2}\int_0^r
y\moy{U_S}^{\ell}_m\mbox{d}y} & =  \\ & -\, \displaystyle{\f{2}{r^2}\int_0^r
y^2\left[\moy{\varepsilon}+\moy{\varepsilon'}\right]^\ell_m\mbox{d}y}\\ \\
 & +\, \displaystyle{\f{2\alpha g}{r^2}\int_0^r
y^2\moy{ \Delta u_z\Delta \theta}^{\ell}_m\mbox{d}y}\\ \\ &   + \, \displaystyle{2\nu\gradr{}{}\moy{(\Delta
    u_i)^2}^{\ell}_m-\f{2\nu\,\ell(\ell+1)}{r^2}\int_0^r\!\moy{(\Delta
  u_i)^2}^{\ell}_m\mbox{d}y} \\ \\ & + \,  
\displaystyle{\moy{NH_\nu}^\ell_m+\moy{NH_p}^\ell_m+\moy{NH_{u_z}}^\ell_m+\moy{NH_{\Delta
      u_z}}^\ell_m}~,  \\ \\
\end{array}
\end{equation}
\begin{equation}
\label{eqTfinal}
\begin{array}{rl}
\displaystyle{\moy{\Theta_R}^\ell_m -\f{\ell(\ell+1)}{r^2}\int_0^r
y\moy{\Theta_S}^{\ell}_m\mbox{d}y} & =  \\ \\
 & -\,  \displaystyle{\f{2}{r^2}\int_0^r
\!\! y^2\left[\moy{N}+\moy{N'}\right]^\ell_m\mbox{d}y}\\ \\
& +\,   \displaystyle{\f{2}{r^2}\int_0^r\!\!\!
  y^2\left(\moy{u_z\diffz{\moy{T}}\Delta\theta}^\ell_m- 
 \moy{u_z'\diffzp{\moy{T'}}\Delta\theta}^\ell_m\right)\mbox{d}y}\\ \\
 & +\,  \displaystyle{2\kappa\gradr{}{}\moy{(\Delta \theta)^2}^\ell_m
-\f{2\kappa\,\ell(\ell+1)}{r^2}\int_0^r\!\moy{(\Delta
  \theta)^2}^\ell_m\mbox{d}y}
\\ \\ & +\, 
\displaystyle{\moy{NH^\theta_\kappa}^\ell_m+\moy{NH^\theta_{u_z}}^\ell_m+\moy{NH^\theta_{\Delta
      u_z}}^\ell_m}~.\\ \\
\end{array}
\end{equation}

\noindent In these equations, the inhomogeneous terms have been separated using
the following notations: 
\begin{eqnarray*}
  \label{defNHall}
\moy{NH_\nu}^\ell_m & = & \f{\nu}{2\,r^2}\int_0^ry^2\diffZZ{}
  \moy{(\Delta u_i)^2}^{\ell}_m\mbox{d}y~, \\
\moy{NH_p}^\ell_m & = & \displaystyle{-\f{2}{\rho\,
  r^2}\int_0^ry^2\diffZ{}\moy{\Delta u_z\Delta p}^{\ell}_m\mbox{d}y}~, \\
\moy{NH_{u_z}}^\ell_m & = & \displaystyle{-\f{1}{r^2}\int_0^r
y^2\diffZ{}\moy{u_z(\Delta u_i)^2}^{\ell}_m\mbox{d}y}~, \\ 
\moy{NH_{\Delta u_z}}^\ell_m & = & \displaystyle{-\f{1}{2\,r^2}\int_0^r
y^2\diffZ{}\moy{\Delta u_z(\Delta u_i)^2}^{\ell}_m\mbox{d}y}
\end{eqnarray*}
and
\begin{eqnarray*}
  \label{defNHthetaall}
\moy{NH^\theta_\kappa}^\ell_m & = & \f{\kappa}{2\,r^2}\int_0^ry^2\diffZZ{}\moy{(\Delta
  \theta)^2}^\ell_m\mbox{d}y~,\\
\moy{NH^\theta_{u_z}}^\ell_m) & = & -\f{1}{r^2}\int_0^r y^2\diffZ{}\moy{u_z
     (\Delta\theta)^2}^\ell_m\mbox{d}y~,\\ 
\moy{NH^\theta_{\Delta u_z}}^\ell_m & = & -\f{1}{2\,r^2}\int_0^r y^2\diffZ{}\moy{\Delta u_z (\Delta
  \theta)^2}^\ell_m\mbox{d}y~.
\end{eqnarray*}
The sums of these inhomogeneous terms in each equation will  be
used later on to characterize the net effect of inhomogeneity. They
will be denoted by

\begin{equation*}
  \moy{NH}^\ell_m=\moy{NH_\nu}^\ell_m+\moy{NH_p}^\ell_m+\moy{NH_{u_z}}^\ell_m+\moy{NH_{\Delta u_z}}^\ell_m~,
\label{defNH}
\end{equation*}
\begin{equation*}
  \moy{NH^{\theta}}^\ell_m=\moy{NH^\theta_\kappa}^\ell_m+\moy{NH^\theta_{u_z}}^\ell_m+\moy{NH^\theta_{\Delta u_z}}^\ell_m~.
\label{defNHtheta}
\end{equation*}
\smallskip

\noindent The usual Kolmogorov equation~(\ref{eqkolmo}) for homogeneous and
isotropic turbulence can be straightforwardly recovered
from equation~(\ref{eqVfinal}). Using incompressibility,
and isotropy, one finds that
$\moy{U_r}=1/(3r^3)\partial_r\left(r^4\moy{(\Delta u_r)^3}\right)$ and 
$\partial_r\moy{(\Delta u_i)^2}=1/r^3\partial_r\left(r^4\partial_r\moy{(\Delta
u_r)^2}\right)$ (\textit{e.~g.} \cite{hill2002}). Furthermore, for
homogeneous and isotropic turbulence (neglecting intermittency effects),
$\moy{\varepsilon}=\moy{\varepsilon'}$ and $\moy{N}=\moy{N'}$ are
constants and 

\begin{equation}
\f{2}{r^2}\int_0^r
y^2\left[\moy{\varepsilon}+\moy{\varepsilon'}\right]^\ell_m\mbox{d}y=\f{4\sqrt{4\pi}}{3}\moy{\varepsilon}r\delta^{\ell
0}\delta_{m0}~,  
\end{equation}

\begin{equation} 
 \displaystyle{\f{2}{r^2}\int_0^r
y^2\left[\moy{N}+\moy{N'}\right]^\ell_m\mbox{d}y}=\f{4\sqrt{4\pi}}{3}\moy{N}r\delta^{\ell0}\delta_{m0}~.
\end{equation}
In the preceding equations, the $\sqrt{4\pi}$ factor comes from the
normalization with respect to $Y^0_0$.
Equation~(\ref{eqkolmo}) follows directly
from integration of equation~(\ref{eqVfinal}) by imposing $\ell=m=0$
(isotropy) and suppressing the $\moy{NH}$ terms (homogeneity). 
As noted by several authors (\textit{e.~g.}
\cite{antonia97}), there is a striking analogy between the generalized
Kolmogorov equation~(\ref{eqVfinal}) which deals with the turbulent 
transport of kinetic energy $(\Delta u_i)^2$ and
equation~(\ref{eqTfinal}), which involves the turbulent transport of
entropy fluctuations, which are proportional to $(\Delta \theta)^2$
for a Boussinesq fluid \citep{lvov91}. This
analogy is perfect when temperature is a passive scalar, while for
non-neutrally stratified flows, extra terms representing the
coupling between temperature and velocity fields  appear.

Finally, it is worth recalling that $\moy{ \Delta u_z\Delta \theta}^{\ell}_m$ stands for
the projection of the scalar correlator $\moy{(\Delta \vec{u}\cdot\vec{e}_z)\Delta \theta}$ on
$Y^m_\ell$. This notation may give the impres\-sion that equation~(\ref{eqVfinal})
is closed in terms of correlation functions all belonging to the same
$(\ell,m)$ order. This is not the case because, as explained in detail
by \cite{biferale03}, gravity breaks the spherical symmetry of the
equations of motion, leading to a coupling between the $\ell$ component of
Kolmogorov equation and the $(\ell-1)$, $\ell$ and $(\ell+1)$
harmonic components of the cross correlator between temperature and
velocity. Therefore, projecting $\moy{\Delta u_z \Delta\theta}$ on
$Y^m_\ell$ only provides a convenient way to analyse the net effect of
gravity on the corresponding $\ell$ component of the third-order 
structure function and does not mean that foliation of the
Rayleigh-B\'enard equations with respect to $\ell$ (that is the question
of the very existence of distinct inertial scaling exponents for 
different $\ell$ components of correlators) is satisfied. Foliation
remains an open question for both Rayleigh-B\'enard and Navier-Stokes
equations \citep{biferalephysrep}.

\subsection{Symmetries of the Boussinesq equations}
Statistical averages in equations~(\ref{eqVfinal})-(\ref{eqTfinal})
obey some symmetry properties which will be listed here before a
detailed analysis is carried out. First, there is an azimuthal
symmetry which is due to isotropy in horizontal planes, so that only
the $m=0$ components need to be considered here. The other important
symmetry is the top-down symmetry of the Boussinesq equations.
For $z=1/2$, the following rules can be derived:

\smallskip

\begin{enumerate}
\item $\moy{U_R}^\ell_0$ vanishes for odd  $\ell$,
\item[]
\item the even $\ell$ components of statistical averages containing 
  odd powers of $u_z$, like for instance $\moy{u_z(\Delta u_i)^2}^\ell_0$,
  vanish, while their $Z$-derivatives are nonzero,
\item[]
\item the $Z$-derivatives of the odd $\ell$ components of the same 
     statistical averages vanish. 
\item[]
\end{enumerate}
Consequently, only even $\ell$ have to be considered at the
center of the cell. Except for the azimuthal symmetry, the previous rules can
not be applied to $z\neq 1/2$, where the complete $\ell$ spectrum
has to be computed. An important consequence of the
previous observations is that inhomogeneous effects should be present at the
center of the convection cell, and that the isotropic component should
be affected by these effects. This also occurs for the
channel flow, which obeys the same top-down symmetry \citep{danaila01}
as Boussinesq convection.

\section{\label{numdetails}Numerics}
The various terms in equations~(\ref{eqVfinal})-(\ref{eqTfinal}) have
to  be computed in order to shed some light on the dominant contributions to 
the third-order structure functions $\moy{\vec{U}}$ and
$\moy{\vec{\Theta}}$ and to test quantitatively the assumptions of
both BO59 and K41 theories for this flow. To this end, one can choose
to resort to either laboratory or numerical experiments.
As will be discussed shortly, the main drawback of numerics is 
that only moderate Rayleigh numbers can be achieved
in comparison to laboratory experiments. However, many interesting
benefits can be gained from using numerical simulations to 
study the properties of
equations~(\ref{eqVfinal})-(\ref{eqTfinal}). The most important ones
are that complete two-points averages can be computed easily without invoking the
Taylor hypothesis and assuming isotropy, and that every quantity such
as $p$ can be easily obtained from the simulation at various locations.
This motivated the choice of using numerics in the present study.
A description of the DNS and numerical procedures used in order to
perform this numerical analysis is now presented.

\subsection{Code setup}
To investigate equations~(\ref{eqVfinal})-(\ref{eqTfinal}), the
Boussinesq equations~(\ref{eqbouss1})-(\ref{eqbouss2})  for
velocity and temperature are solved for $\ray=10^6$, $\pr=1$ and an
aspect ratio $\aspra=5$ with a DNS code
initiated by \cite{cali}, designed for anisotropic flows. It
makes use of a sixth-order compact finite differences scheme
\citep{lele92} in the  vertical $z$ direction  and of a spectral
scheme with 2/3 dealiasing in the horizontal ones. Time-stepping is
implemented via a third-order, fully explicit Runge-Kutta scheme
\citep*{demuren01}.  A projection method is used to ensure that the
velocity field remains divergence-free. This method involves a
Poisson solver to compute the pressure field.
For the chosen aspect ratio, $256$ points are used in each horizontal direction and $128$
points in the vertical. A smaller mesh size is required in the
vertical to resolve the thermal boundary layers accurately. 
Fixed temperature and stress-free boundary conditions are imposed on
both plates, while periodic boundary conditions are used in the
horizontal directions. Note that these boundary conditions offer
a higher supercriticality  (for a given Rayleigh number) than  rigid
boundaries, so that this simulation at $\ray=10^6$ is in
all likelihood located at the lower end (with respect to $\ray$) of the
soft turbulence regime range of convection \citep*{heslot87}.

Distances are normalized with respect to $d$:
$z=0$ corresponds to the bottom plate and $z=1$ to the top one.
The simulation starts from small random temperature perturbations of
the initial linear temperature profile and 
lasts for approximately fifty turnover times of the
fluid, where one turnover time, defined here as twice the time it
would take for a fluid element with velocity equal to the
r.m.s. vertical velocity in the central plane to travel from one boundary
to the other, corresponds to the time interval required before two
snapshots can be considered independent. After a short linear growth phase, an
energetically steady turbulent state is reached in approximately one
turnover time. It should be noted that plume
clustering \citep{cattaneo01,hartlep03,parodi04,rincon05} is observed
in moderate and large aspect ratio simulations, which can be identified with
the mean wind observed in small aspect ratio convection cells, as shown
recently by \cite{reeuwijk05}. The slow
evolution of the integral scale towards large scales that results from
this effect does not affect the results presented here, which concern
structures on scales smaller than the layer depth, for which an
energetically steady state is reached  quickly. In the
statistically steady state, the Nusselt number computed from the data
of the simulation is $\nuss\sim 14$. The Bolgiano scale, following
equation~(\ref{bolglength}), is $L_B=0.11$ (with respect to $d$), 
but as will be shown later, this value does not correspond to the
effective Bolgiano scale at the center of the convection cell. The
Kolmogorov dissipation scale, $\eta=0.016$, is adequately resolved
with the chosen mesh sizes. At the center of the convection cell, the
Reynolds number based on the r.m.s. velocity and the layer depth is
$\rey\simeq 320$ and the Taylor Reynolds number is $\rey_\lambda\simeq 30$.

\subsection{Averaging procedure and statistical convergence}
Statistical averages are computed for given horizontal planes $z$
by measuring each required quantity once every turnover time and at every
four points in each horizontal direction. This procedure
has notably been used by \cite{arad99b} for the channel flow and makes
uses of homogeneity in both horizontal directions. 
At this stage, one may wonder why only a moderately turbulent simulation
with $\ray=10^6$ is presented here, whereas \cite{verzicco03}
managed to simulate convective flows up to $\ray=2\times 10^{11}$.
It is worth emphasizing that the present work aims at computing
various two-point statistical averages, and that such computations
require a very long integration of the equations and a large number of
independent realisations of the flow to be collected in order to get
a good convergence, especially for structure functions of odd order and
large $\ell$ components. 
This very large data and computing time requirement is the main
reason why all numerical studies that deal with the computation of
statistical quantities such as structure functions have to make an
important compromise between resolution and integration
time and are therefore currently limited to moderate Rayleigh/Reynolds
numbers. For instance, \cite{arad99b} used 160 independent snapshots of
their flow at $\rey_\lambda=70$ to compute structure functions, while
 \cite{casciola03} simulated shear flow turbulence at $\rey_\lambda=45$
 during 5900 shearing times in order to perform their statistical analysis.
\cite{calzavarini02} and \cite{biferale03} stored respectively 400 and
270  independent configurations of aspect ratio one convective
turbulence at Rayleigh numbers of order $10^7$.
Such parameter values all lead to turbulent regimes comparable to that of
this simulation (in the case of \cite{casciola03}, the largest value for $r$
is $55\eta$, to be compared with $30\eta$ here). In the present study,
convergence for all statistical quantities at correlation lengths
smaller than 0.5 starts to be satisfactory (up to a few percents) after
approximately ten turnover times,
which is only possible because one time snapshot of a simulation with
aspect ratio $A=5$ can be considered ideally as 25 independent snapshots
of an aspect ratio one simulation for scales smaller than $0.5$.
Therefore, convergence is in fact only obtained by collecting more than 200
independent ``aspect ratio one'' snapshots, as in the previously mentioned  studies.
In comparison, the high-resolution simulations by \cite{verzicco03} with
aspect ratio $A=1/2$ last for at most 220 free-fall turnover times,
which correspond to only 16 turnover times as defined here\footnote{Snapshots
  separated by one free-fall turnover time in their study can not be
  considered independent. The free-fall turnover time in their
  simulation is approximately 14 times smaller than the
turnover time based on the r.m.s. vertical velocity, since the latter is
approximately 0.07 times the free-fall velocity (see their figure~14).}. 
This is far from sufficient to obtain convergence for all
statistical quantities of interest and, unfortunately, performing such
simulations for much longer times is currently prohibitive from the
computing time point of view.

\subsection{SO(3) decomposition}
The spherical harmonics decomposition is computed in the following
way: statistical averages are  computed for various
correlation length vectors $\vec{r}$ making different angles $\theta$ with the
vertical direction, located on a Gauss-Legendre grid containing 15
points. This computation is performed for 20 different values of
the azimuthal angle $\varphi$. A Fourier transform on $\varphi$ and a
Legendre transform on $\theta$ follow, which lead to the amplitudes
of the various $(\ell,m)$ components at a given $r$. The procedure is
repeated at various distances $r$ to obtain the final results.
Interpolation between grid points  is the only
trick that one has to resort to in order to compute anisotropic structure
functions. It does not affect the results significantly except
for very small correlation lengths comparable to the mesh size. 
The $\partial_Z$ and $\partial^2_Z$ derivatives in 
the inhomogeneous terms are calculated by differentiating results
obtained at different $z$ with fourth-order finite difference schemes.

\section{\label{results}Results}
This section provides an exhaustive presentation of  results
obtained by computing the various terms in
equations~(\ref{eqVfinal})-(\ref{eqTfinal}) from DNS data. A
preliminary analysis  is first carried out in order to outline 
inhomogeneous effects in the Rayleigh-B\'enard system.

\subsection{Inhomogeneity in turbulent Rayleigh-B\'enard convection}
\begin{figure}
   \centering
 \includegraphics[width=6.5cm]{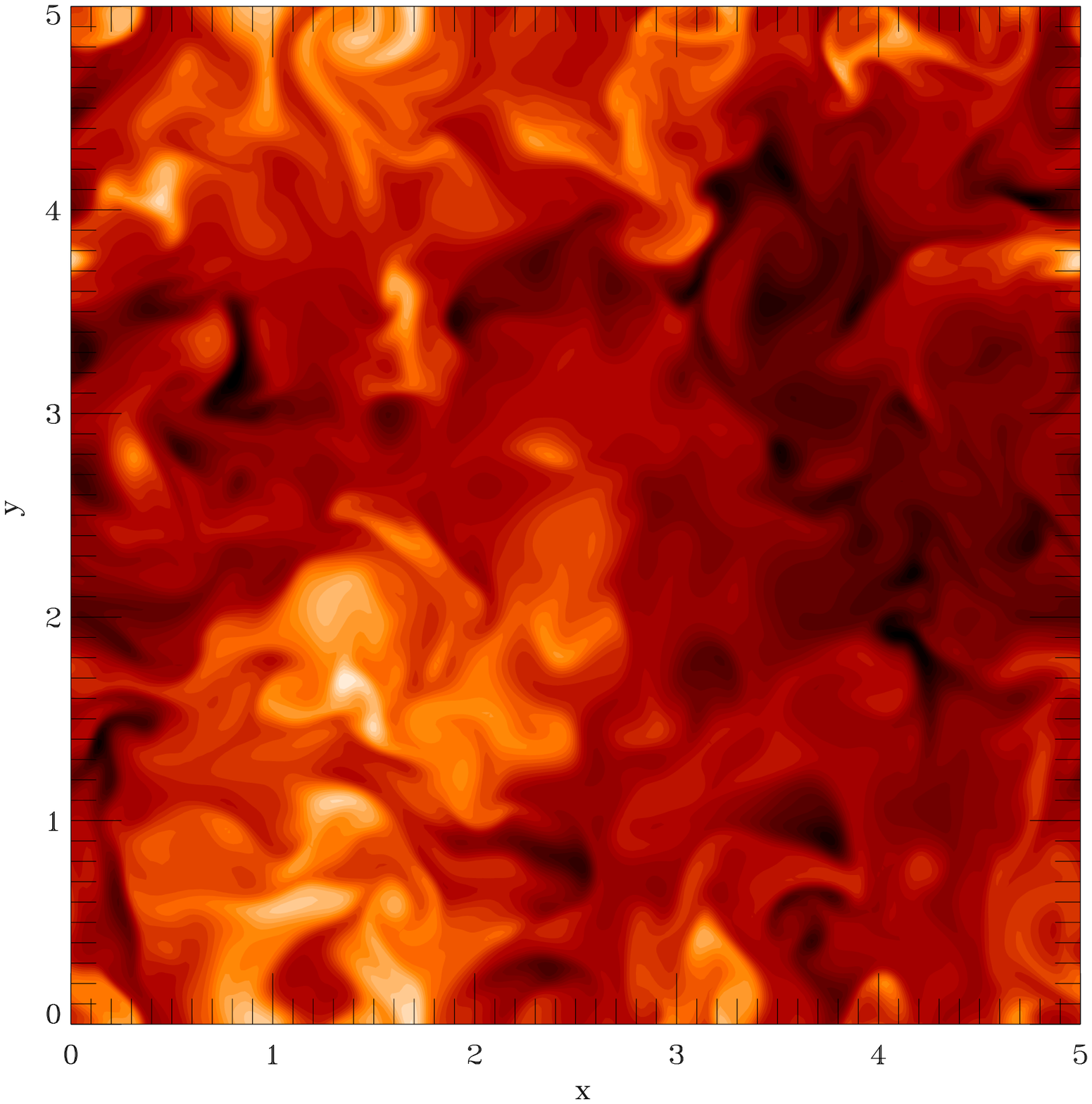}\includegraphics[width=6.5cm]{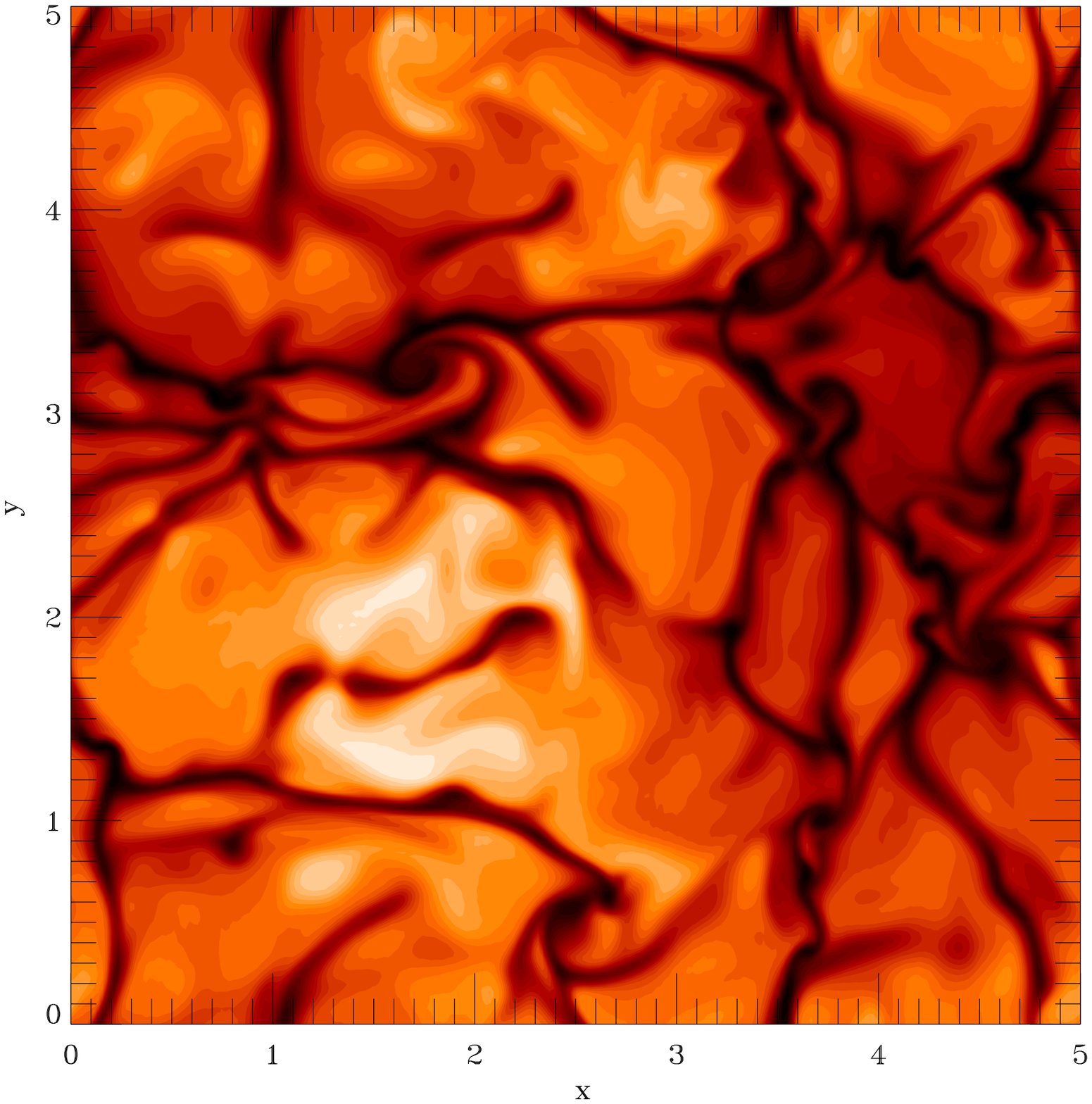}
   \caption{Horizontal maps of temperature fluctuations at $z=0.5$ (left) and
     $z=0.9$ (right).}
   \label{fig1}
 \end{figure}

\begin{figure}
  \centering
\includegraphics[height=7.cm]{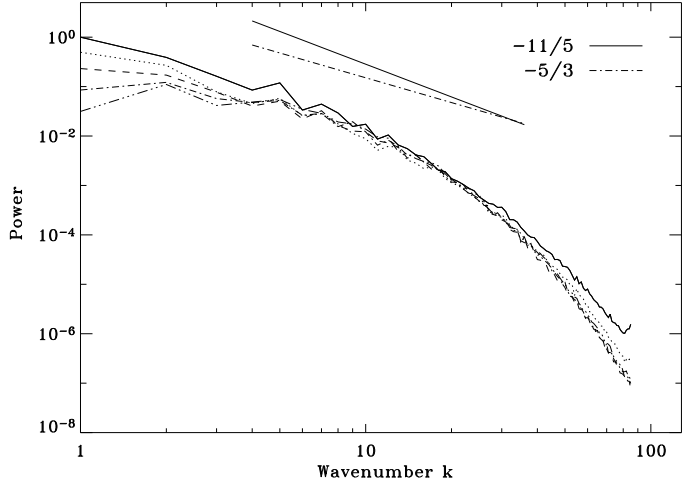}
  \caption{Velocity spectra defined in
    equation~(\ref{defspectre}) for $z=0$  (\fullline), 
    $z=0.13$ (\dotline), $z=0.25$ (\dashdotline), $z=0.38$
    (\dashtripledotline). $z=0$ corresponds to the bottom of the
    convective cell. Due to the top-down symmetry of Boussinesq
    convection, only spectra in the lower half of the layer are
    presented. BO59 and K41 slopes have been overplotted to illustrate
    the difficulty to identify a definite spectral slope. Wave numbers
    have been normalized with respect to $2\pi/(dA)$, so that $k=1$
    corresponds to the largest horizontal scale of the domain.}
  \label{fig2}
\end{figure}

Figure~\ref{fig1} represents snapshots of temperature perturbations in the
midlayer and close to the surface in the statistically  stationary
state of the simulation. Obviously, temperature maps at different
altitudes look very different: while large scale temperature
perturbations  dominate the convective pattern inside the cell,
smaller scale structures separated by dark lanes become prominent in the
thermal boundary layer close to the surface.  The presence of these dark lanes
also leads to a clear increase of small-scale power in the temperature spectra
computed close the plates \citep{rincon05}. This
depth-variation of the properties of convective turbulence has been known for a
long time. It is even more striking in non-Boussinesq compressible
convection, where the top-down symmetry is lost (\textit{e. g.}
\cite{toomre90,cattaneo91}). 
Inhomogeneity in the system is
further illustrated in figure~\ref{fig2}, which
shows horizontal velocity spectra computed at various altitudes
$z$. These spectra are defined according to
  \begin{equation}
\label{defspectre}
E(k,z)=\int_{\Omega_{\vec{k}}}|\hat{\vec{u}}^{\phantom{*}}\!\!\null_{\vec{k}}(z)|^2k\,
\mbox{d}\Omega_{\vec{k}}~,
\end{equation}
where $\Omega_{\vec{k}}$ stands for the wave vectors angles in the
horizontal spectral plane, and the hat denotes horizontal Fourier transforms.
Measuring definite slopes on these spectra proves
all the more difficult than their shape depend quite
strongly on the depth at which the computation is carried
out. As will be demonstrated  in the next paragraphs,
there are various reasons for this absence of scaling laws. Note
finally that the previously mentioned small-scale structures that appear on
temperature maps close to the surface are essentially temperature
perturbations, which explains why they do not show up in the
corresponding velocity spectra at intermediate wave numbers.

\subsection{Isotropic component of  equations~(\ref{eqVfinal})-(\ref{eqTfinal})}
The first step in the analysis of the simulation is the examination of
the isotropic parts of equations~(\ref{eqVfinal})-(\ref{eqTfinal}), which
take on the following form: 
 \begin{align}
- \moy{U_R}^0_0+\f{2\alpha g}{r^2}\int_0^r
y^2\moy{ \Delta u_z\Delta \theta}^0_0\mbox{d}y+2\nu\gradr{}{}\moy{(\Delta
    u_i)^2}^0_0+\moy{NH}^0_0=\qquad\qquad\qquad\notag\\  \f{2}{r^2}\int_0^r
y^2\left[\moy{\varepsilon}+\moy{\varepsilon'}\right]^0_0\mbox{d}y~,
\label{eqVfinal_l=0}
 \end{align}
 \begin{align}
   \label{eqTfinal_l=0}
-\moy{\Theta_R}^0_0  + \f{2}{r^2}\int_0^r
  y^2\left(\moy{u_z\diffz{\moy{T}}\Delta\theta}^0_0- 
 \moy{u_z'\diffzp{\moy{T'}}\Delta\theta}^0_0\right)\mbox{d}y
\qquad\qquad\qquad\qquad \notag \\ 
+2\kappa\gradr{}{}\moy{(\Delta \theta)^2}^0_0
+\moy{NH\,^\theta}^0_0 = \qquad\qquad\notag\\ 
\f{2}{r^2}\int_0^r y^2\left[\moy{N}+\moy{N'}\right]^0_0\mbox{d}y
\end{align}
In the following, the right hand sides of
equations~(\ref{eqVfinal_l=0})-(\ref{eqTfinal_l=0}) will be shown to be always
close to $4\sqrt{4\pi}/3\moy{\varepsilon} r$ and
$4\sqrt{4\pi}/3\moy{N} r$ respectively, except on large scales $r$ which
correspond to penetration of the correlation length vector into the
boundary layers, where dissipation (and so $\moy{\varepsilon'}$)
increases. For clarity reasons, these terms will thus be referred to
as ``$4/3\moy{\varepsilon} r$'' and ``$4/3\moy{N} r$'' terms.

\subsubsection{Central plane analysis}
\begin{figure}
\centering
\hspace{-0.75cm}\includegraphics[width=7.3cm]{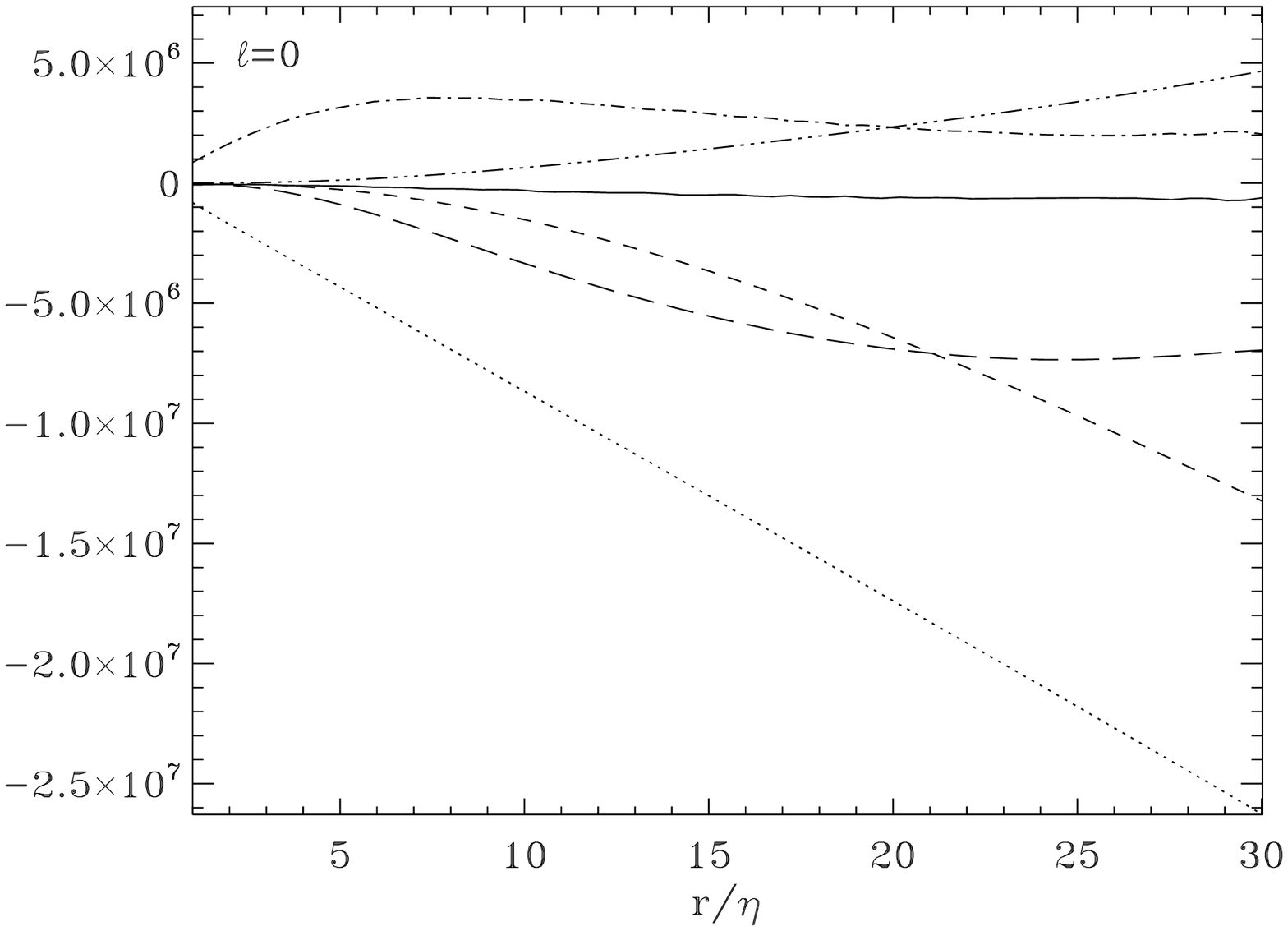}\hspace{-0.8cm}
\includegraphics[width=7.3cm]{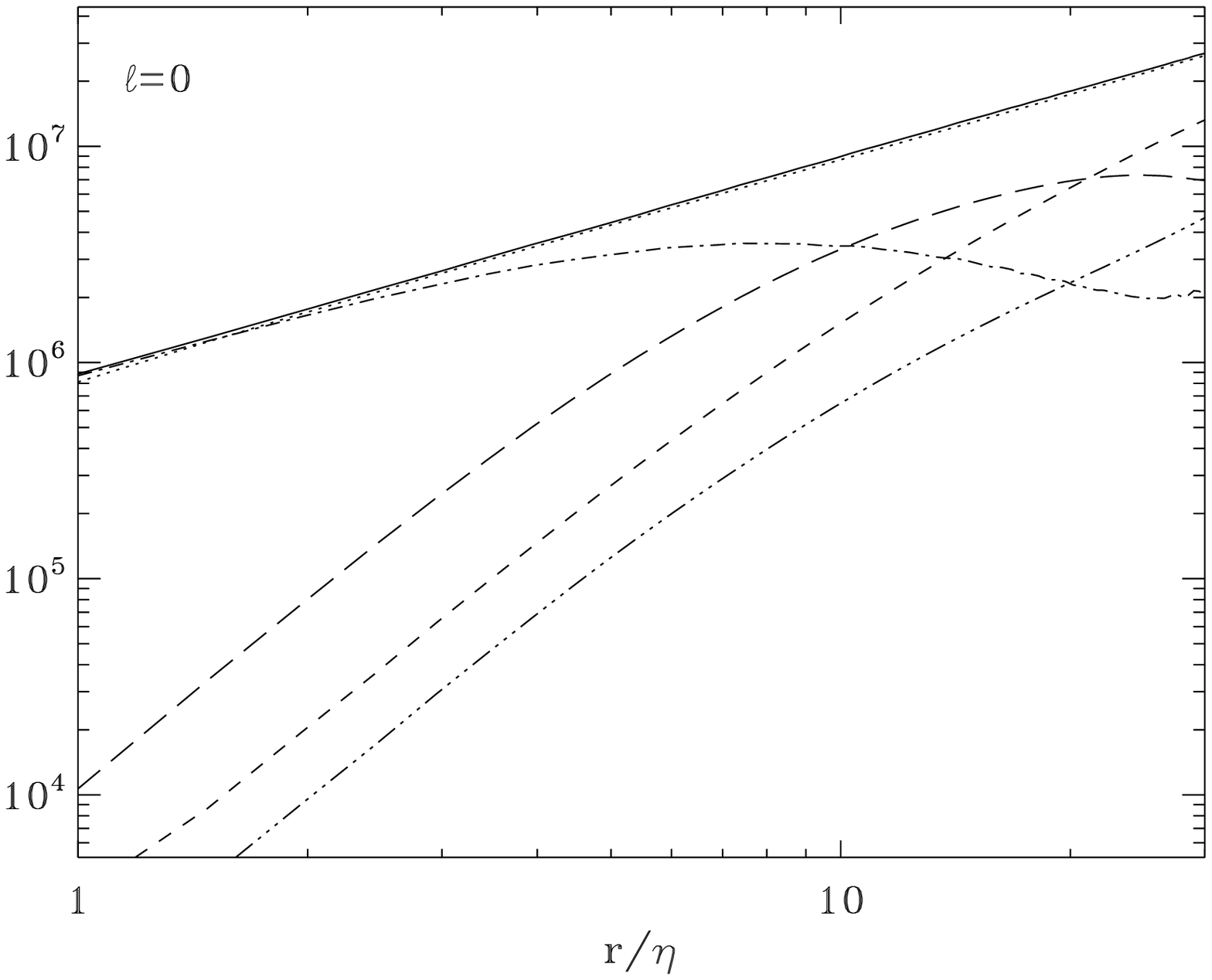}\hspace{-0.8cm}
  \caption{Two different representations of the scale-by-scale budget
   equation~(\ref{eqVfinal_l=0}). Left: linear-linear plots of
   $\moy{U_R}^0_0$ (\longdashline), 
$-2\alpha g/r^2\int_0^r y^2\moy{ \Delta u_z\Delta \theta}^0_0\mbox{d}y$
   (\dashline), $2\nu\partial_r\moy{(\Delta u_i)^2}^0_0$
   (\dashdotline), $\moy{NH}^0_0$ (\dashtripledotline),
 $-4/3\moy{\varepsilon}r$ term (\dotline) and net budget (\fullline).
 Right: log-log plots of  $-\moy{U_R}^0_0$ (\longdashline), 
$2\alpha g/r^2\int_0^r y^2\moy{ \Delta u_z\Delta \theta}^0_0\mbox{d}y$
   (\dashline), $2\nu\partial_r\moy{(\Delta u_i)^2}^0_0$
   (\dashdotline), $\moy{NH}^0_0$ (\dashtripledotline). The sum of the
   l.h.s. terms of equation~(\ref{eqVfinal_l=0})  (\fullline) is seen
   to be in very good balance with the r.h.s. $4/3\moy{\varepsilon}r$ term
   (\dotline).}
  \label{figV_l=0}
\end{figure}

Focus is first given to the central plane $z=1/2$.
An illustration of equation~(\ref{eqVfinal_l=0}) at $z=1/2$ is
given in figure~\ref{figV_l=0}. The right hand side of
equation~(\ref{eqVfinal_l=0}) is almost linear with a 
slope $4\sqrt{4\pi}/3\moy{\varepsilon}$. This $4/3\moy{\varepsilon} r$ term is the
dominant contribution to the third-order structure function at
moderate to large scales, but  forcing by buoyancy is also important in
the same range, as seen on the  left plot. As shown by
\cite{calzavarini02}, the estimate of the Bolgiano length
(\ref{bolglength}) is rather crude and the
effective Bolgiano length (corresponding to the crossover between the
buoyancy term and $-4/3\moy{\varepsilon} r$) 
at the center of the convection cell which can be extrapolated from
figure~\ref{figV_l=0} is more likely closer to 1 ($r/\eta\simeq 60$)
than to 0.1 ($r/\eta=6$). 
The aspect ratio and set of boundary conditions have a direct
influence on the prefactors of the various
relations leading to equation~(\ref{bolglength}) and thus
affect the value of the effective Bolgiano length. Another simulation
(not presented here) with aspect ratio 1 and all other parameters remaining
constant notably reveals that $L_B$ tends to get smaller in the central plane
for convection cells with smaller aspect ratio. This effect may
be related to the presence of a mean wind (plume clustering): as shown by
\cite{rincon05}, the scale of the mean wind corresponds to the scale
at which a maximum power is injected through buoyancy. In an aspect
ratio 1 simulation, these structures have to concentrate in a cube of
aspect  ratio 1, while they are allowed to spread over larger horizontal scales at
larger aspect ratio. Buoyancy forcing may consequently be more important
at smaller scales in small aspect ratio convection, thus reducing the
effective Bolgiano length. Note finally that equation~(\ref{bolglength})
does not take into account the fact that the Bolgiano length derived by
Bolgiano and Obukhov depends on $z$-dependent dissipation rates and is
therefore $z$-dependent as well. Using this exact expression, as done
notably by \cite{calzavarini02},  does not lead to
order of magnitude changes for $L_B$ in the bulk of the convection cell
in comparison to the estimation provided by equation~(\ref{bolglength})
for the present simulation. In fact, both expressions rely on dimensional
analysis and must therefore be understood as estimations up to order one
prefactors.

\begin{figure}
  \centering
 \includegraphics[height=7cm]{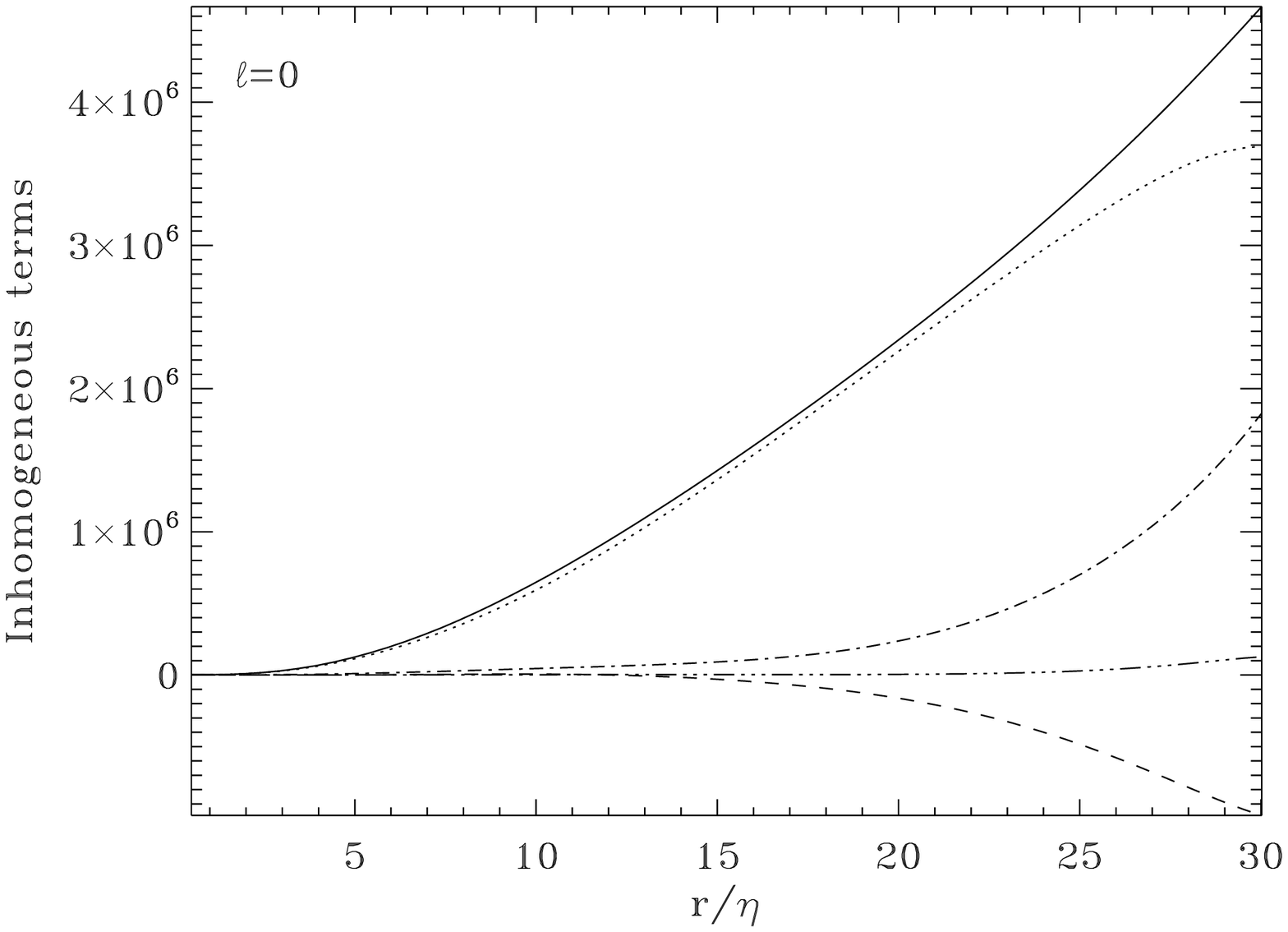}
  \caption{Decomposition of $\moy{NH}^0_0$ (\fullline) into
$\moy{NH_\nu}^0_0$ (\dashtripledotline),
$\moy{NH_p}^0_0$ (\dashdotline),
$\moy{NH_{u_z}}^0_0$ (\dotline),
$\moy{NH_{\Delta u_z}}^0_0$ (\dashline).
 The large scale inhomogeneity $\moy{NH_{u_z}}^0_0$ dominates the
    overall inhomogeneous contribution to equation~(\ref{eqVfinal_l=0}).}
  \label{decNHl=0}
\end{figure}

Using the effective value $L_B\simeq 1$ for this simulation, one
can estimate that $L_B/\eta\simeq 60$, so that there should in
principle be some room for K41 scalings to occur for  $\eta<
r<L_B$. However, the buoyancy term  always remains of the
same order of magnitude as the $-4/3\moy{\varepsilon} r$ term. 
Also, these two terms have opposite sign, which leads to a significant
modification of the third-order structure function in comparison to 
isotropic turbulence forced on the largest scales only. 
The inhomogeneous term $\moy{NH}^0_0$ also gives a positive contribution that
compensates the $-4/3\moy{\varepsilon} r$ term. This inhomogeneous
term increases with $r$ and becomes larger than the viscous term for
$r\simeq 20 \eta$. A detailed analysis of
this term (figure~\ref{decNHl=0}) shows that the main
contribution comes from $\moy{NH_{u_z}}^0_0$. The other inhomogeneous
terms have a smaller influence on the global budget of
equation~(\ref{eqVfinal_l=0}).
\cite{danaila01} have shown that the $\moy{NH_{u_z}}$ term, which
they have identified as a large-scale
inhomogeneity, is also the dominant inhomogeneous contribution in 
channel flow turbulence.

\begin{figure}
  \centering
  \includegraphics[height=7cm]{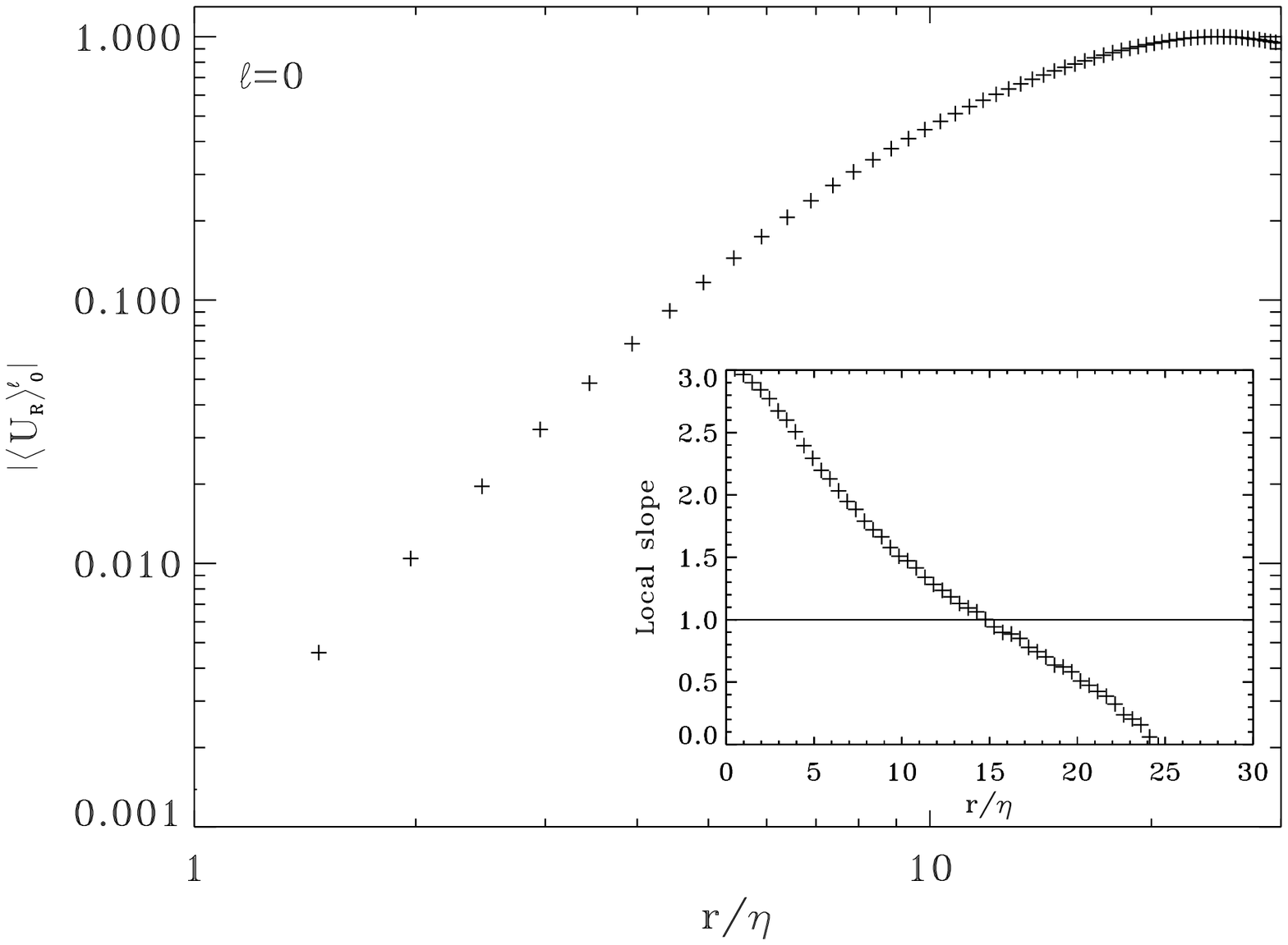}
  \caption{$|\moy{U_R}^0_0|$ as a function of $r/\eta$, normalized
    with respect to its maximum over $r$. 
    Inset: logarithmic slope of the same function. A K41
    scaling would lead to  $\moy{U_R}^0_0\sim r$ (up to a sign change).}
  \label{figS3l=0}
\end{figure}

The most important consequence of having several important contributions to 
the third-order structure function is that it does not exhibit any
scaling behaviour. To illustrate this, a plot of $\moy{U_R}^0_0$ is given in
figure~\ref{figS3l=0} together with a plot of the associated local
logarithmic slope. It proves impossible to identify an inertial range on
this plot, but note as a verification that the local slope of $\moy{U_R}^0_0$ tends to 3 at
very small scales, as expected from classical arguments. It is often
claimed that the absence of scaling behaviour in various flows is due
to the smallness of the Reynolds number, because in that case the viscous term in
equation~(\ref{eqkolmo}) can not be neglected a most scales.
This is only partly true for this convective flow,
since the viscous term appears to be far smaller than the $-4/3\moy{\varepsilon}
r$ term for $r>0.15$ ($r/\eta>10$). Strictly speaking, the smallness
of the Reynolds number is mainly responsible for the absence of a very
large scale separation between $L_B$ and $\eta$ here, which leads to
the previously mentioned compensation effect. As will be discussed in
\S\ref{discuss}, definite scalings in the range $\eta<r<L_B$ may only be
observed for very large Rayleigh numbers.
In any case, in a homogeneous flow forced on large scales
only, for which no compensation by a forcing mechanism exist, 
scalings should in principle be observed, at least in a restricted
subrange. There is however a supplementary effect which may prevent
scaling laws from showing up even in this favourable situation,
namely anisotropy. In that case, the occurrence of scalings may depend
on the way data is analysed. This point will be further investigated
later on in the paper.

Comparing convective turbulence to other types of flows such as
channel flows and shear flows shows that 
many features of turbulence are shared by all these flows, even though their
forcing mechanisms are completely different. 
For instance, the right plot of figure~\ref{figV_l=0} can easily be compared to
 the figure~5 of \cite{danaila01}.
These authors studied the effect of inhomogeneity on third-order
structure functions for the channel flow by assuming  isotropy,
which is however not completely satisfied in the flow. The main difference
with the channel flow is that the $\moy{NH}^0_0$ term  does not tend
to balance $-4/3\moy{\varepsilon}r$ directly in turbulent convection at the
largest scales that can be analysed using the SO(3) technique, since
buoyancy forcing becomes increasingly important at such
scales. Instead, the inhomogeneous term tends to compensate the net
effect of $-4/3\moy{\varepsilon}r$ and of buoyancy (note that the
situation is slightly different from the previously mentioned channel flow
study, since the SO(3) analysis is restricted to scales smaller than
the integral scale of the flow, so that $\moy{U_R}^0_0$ does not tend
to zero yet at the largest scales available).
Another interesting comparison  with the results of \cite{casciola03}
for homogeneous shear flow (their figure~1) and the left plot of
figure~\ref{figV_l=0} can be made. Their results are
restricted to $\ell=0$, since their scale-by-scale budget results from
averaging structure functions over a $\vec{r}$-sphere.
Except for inhomogeneous effects, the scale-by-scale budgets of the
two flows are very similar. Buoyancy appears to play almost
the same role in Rayleigh-B\'enard convection as shear forcing in a shear flow.
The same conclusions regarding the determination of structure
functions exponents can be drawn for both flows, since the forcing
remains important at all scales in both cases. 

\begin{figure}
\centering
\hspace{-1.4cm}
\includegraphics[width=7.4cm]{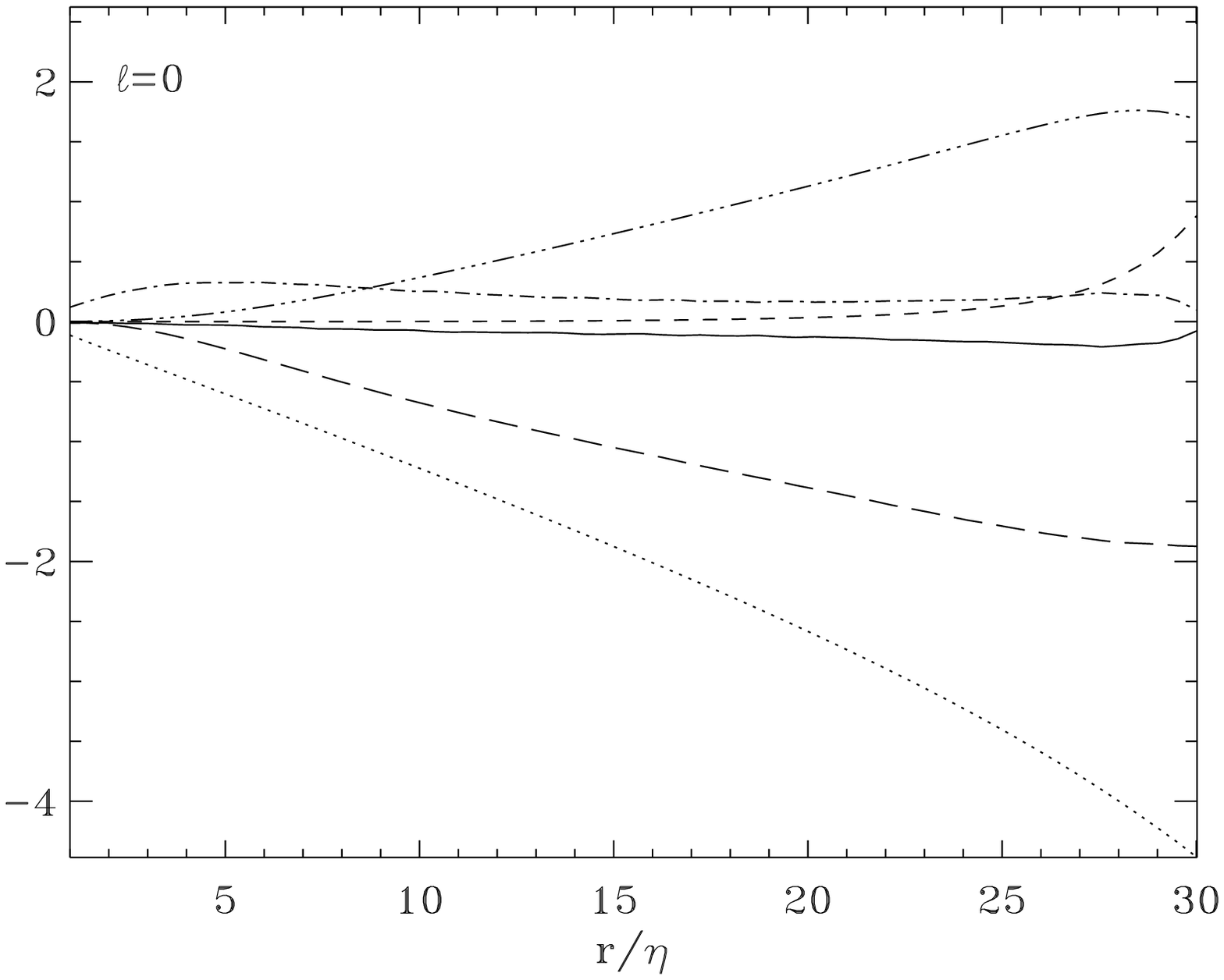}\hspace{-0.8cm}
\includegraphics[width=7.4cm]{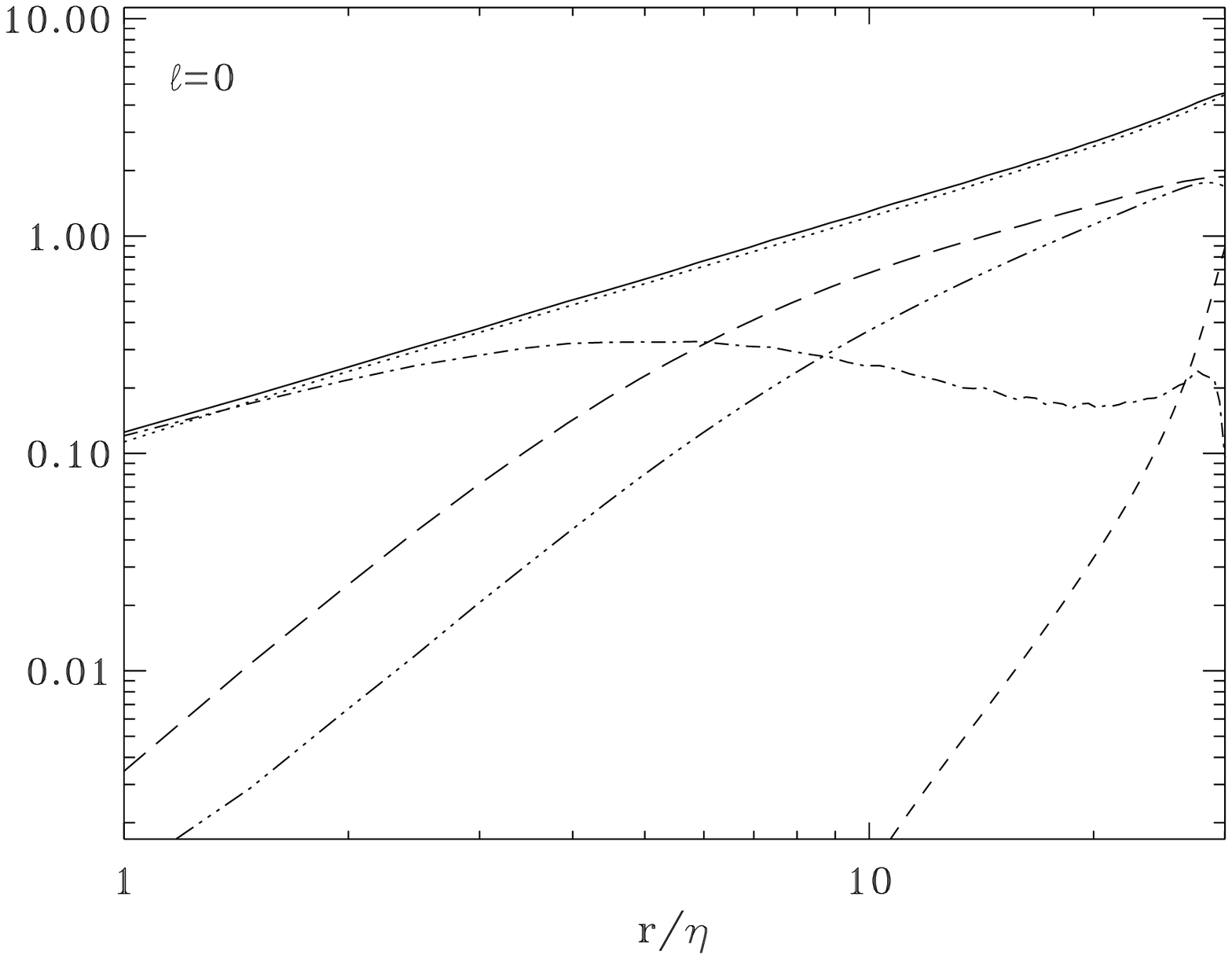}\hspace{-0.8cm}
  \caption{Representations of the temperature scale-by-scale budget
   governed by equation~(\ref{eqTfinal_l=0}). Left: linear-linear plots
   of $\moy{\Theta_R}^0_0$ (\longdashline), $2/r^2\int_0^r
  y^2(\moy{u_z\diffz{\moy{T}}\Delta\theta}^0_0- 
 \moy{u_z'\diffzp{\moy{T'}}\Delta\theta}^0_0)\mbox{d}y$ (\dashline),
$2\kappa\partial_r\moy{(\Delta \theta)^2}^0_0$ (\dashdotline),
$\moy{NH\,^\theta}^0_0$ (\dashtripledotline), $-4/3\moy{N}r$ term
   (\dotline) and net budget (\fullline). 
Right: log-log plots of $-\moy{\Theta_R}^0_0$ (\longdashline), $2/r^2\int_0^r
  y^2(\moy{u_z\diffz{\moy{T}}\Delta\theta}^0_0- 
 \moy{u_z'\diffzp{\moy{T'}}\Delta\theta}^0_0)\mbox{d}y$ (\dashline),
$2\kappa\partial_r\moy{(\Delta \theta)^2}^0_0$ (\dashdotline),
$\moy{NH\,^\theta}^0_0$ (\dashtripledotline).
    The sum of the l.h.s. terms of equation~(\ref{eqTfinal_l=0})
    (\fullline) is seen to be in very good balance with the
    r.h.s. $4/3\moy{N}r$ term (\dotline).}
  \label{figT_l=0}
\end{figure}

\begin{figure}
\centering
  \includegraphics[height=7.cm]{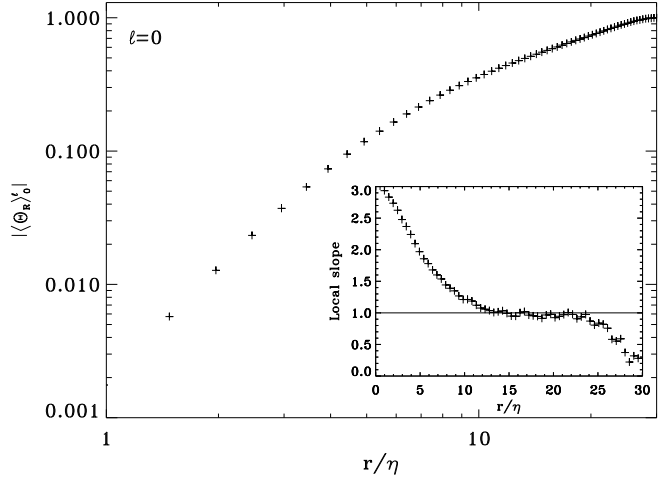}
  \caption{$|\moy{\Theta_R}^0_0|$ normalized to its maximum over $r$ 
  and its local logarithmic slope (inset), as a function of $r/\eta$.}
\label{slopeTheta}
\end{figure}

\begin{figure}
  \centering
 \includegraphics[height=7cm]{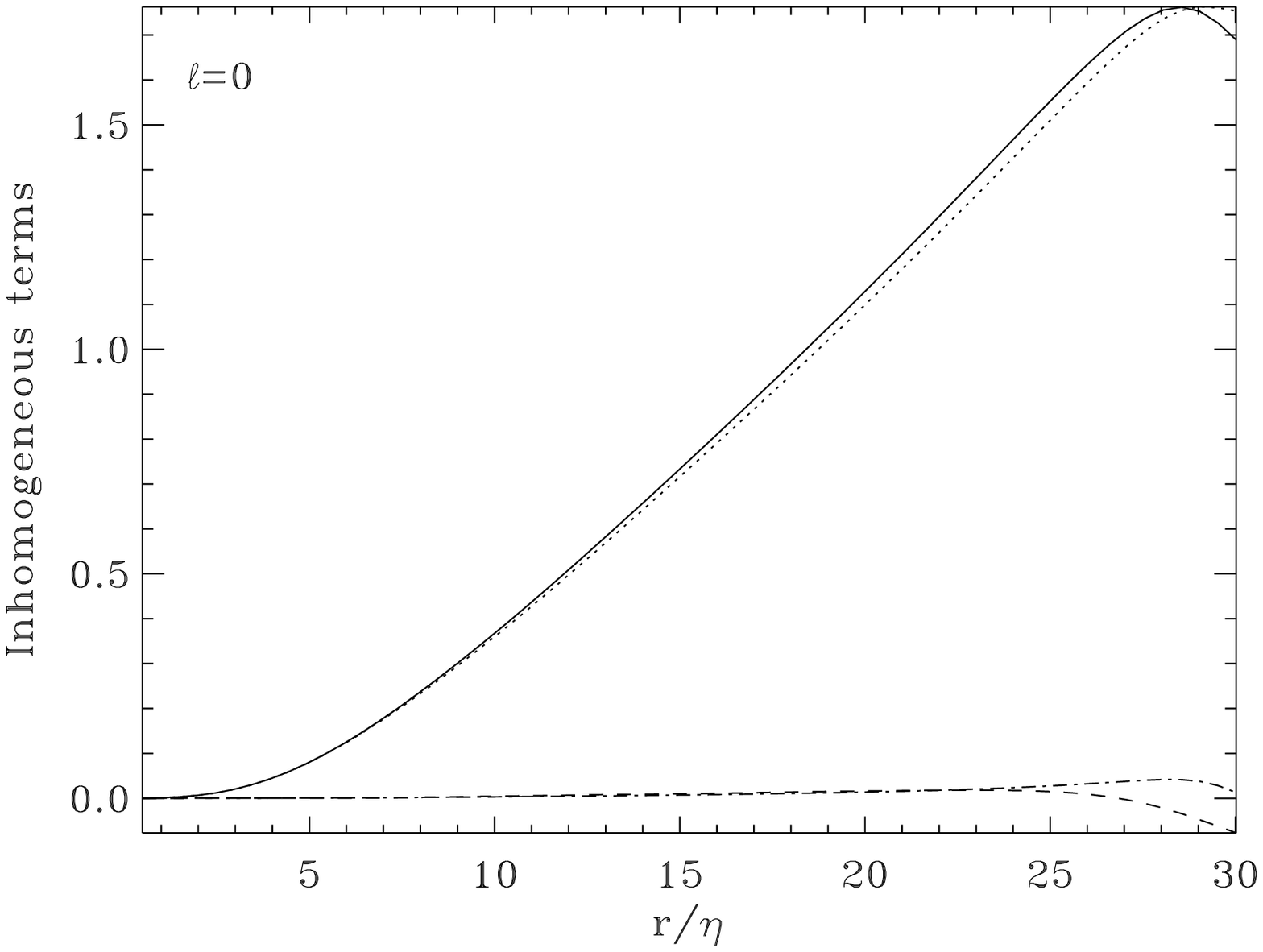}
  \caption{Decomposition of $\moy{NH^\theta}^0_0$ (\fullline) into 
$\moy{NH^\theta_\kappa}^0_0$ (\dashdotline),
$\moy{NH^\theta_{u_z}}^0_0$ (\dotline),
$\moy{NH^\theta_{\Delta u_z}}^0_0$ (\dashline).
 As for the velocity  equation, the large-scale inhomogeneity
 $\moy{NH^\theta_{u_z}}^0_0$ dominates the
 overall inhomogeneous contribution to equation~(\ref{eqTfinal_l=0}).}
  \label{decNHTl=0}
\end{figure}

Equation~(\ref{eqTfinal_l=0}) for temperature fluctuations is also
worth analysing in the central plane. Figure~\ref{figT_l=0} represents
the various terms in this equation in a similar way as was done in
figure~\ref{figV_l=0}.
A plot of $\moy{\Theta_R}^0_0$
is presented in figure~\ref{slopeTheta}, together
with the  associated logarithmic local slope. A clear
plateau close to value 1 is observed for the logarithmic slope for
 $12\eta<r<25\eta$. This plateau corresponds to a dominant balance
between  $\moy{\Theta_R}^0_0$ and the $-4/3\moy{N}r$ term, which is required
in both BO59 $\left(r\sim\left(r^{3/5}\times\left(r^{1/5}\right)^2\right)\right)$
and K41 $\left(r\sim\left(r^{1/3}\times\left(r^{1/3}\right)^2\right)\right)$ theories.  
Figure~\ref{figT_l=0} shows notably that the BO59 hypothesis which
neglects advection of the mean temperature profile term in comparison to
the $-4/3\moy{N}r$ term is very well satisfied in the simulation,
because the core of the convection cell is almost
isothermal at $\ray=10^6$, so that the $\diffz{}\moy{T}$ terms are
negligible in this region. Inhomogeneity is seen to give a significant
contribution to the third-order structure function, mainly through the action of
large-scale inhomogeneities $\moy{NH^\theta_{u_z}}^0_0$
(figure~\ref{decNHTl=0}), as in the velocity equation.
However, this does not prevent scaling behaviour in a small
``inertial'' range in that case. The reason for this is that 
the dominant inhomogeneous term $\moy{NH^\theta_{u_z}}^0_0$, as can be
seen on the same plot, is also proportional to $r$ on large scales
(right plot of figure~\ref{figT_l=0}).
Note that the observation of scalings for temperature related
structure functions but not for velocity structure
functions can not simply be attributed to the present value of the
Prandtl number, because the potential scaling ranges of $\moy{U_R}^0_0$ and
$\moy{\Theta_R}^0_0$  should be similar for $\pr=1$. This analysis
therefore confirms that moderate values of the Taylor Reynolds or
Peclet numbers (the equivalent of the Reynolds number for
thermal diffusion) do not necessarily imply that an inertial range
should not exist. 

\subsubsection{Analysis off the central plane}
The main problem with the SO(3) decomposition of structure functions
is that the analysis must be restricted to rather small scales $r$ in
regions close to the boundaries of the system (for wall-type boundary
conditions). In such
regions, the computation of reduced structure functions with $\vec{r}$
lying in planes parallel to the boundaries
remains the only available tool
(see \cite{calzavarini02} for an application to
Rayleigh-B\'enard convection). 
Remembering that the symmetry properties listed in
\S\ref{theorie} disappear off the central plane, it is however
interesting to push the SO(3) analysis as far as possible from this
region in order to observe trends of the $z$-dependence of
inhomogeneous effects. A  brief description of turbulent
statistics at $z=1/4$ is given to this end. At this altitude, the maximum $r$ for
which turbulent statistics can be computed is only $15\eta$, so that
care must be taken in the interpretation of the results. For such a
restricted range of correlation lengths, it is clear that at most a
beginning of inertial range can be expected and that probing
inhomogeneous effects proves difficult, since these effects
should be dominant on scales comparable to the integral scale of the
flow only.

\begin{figure}
\centering
\hspace{-1.cm}\includegraphics[width=7.3cm]{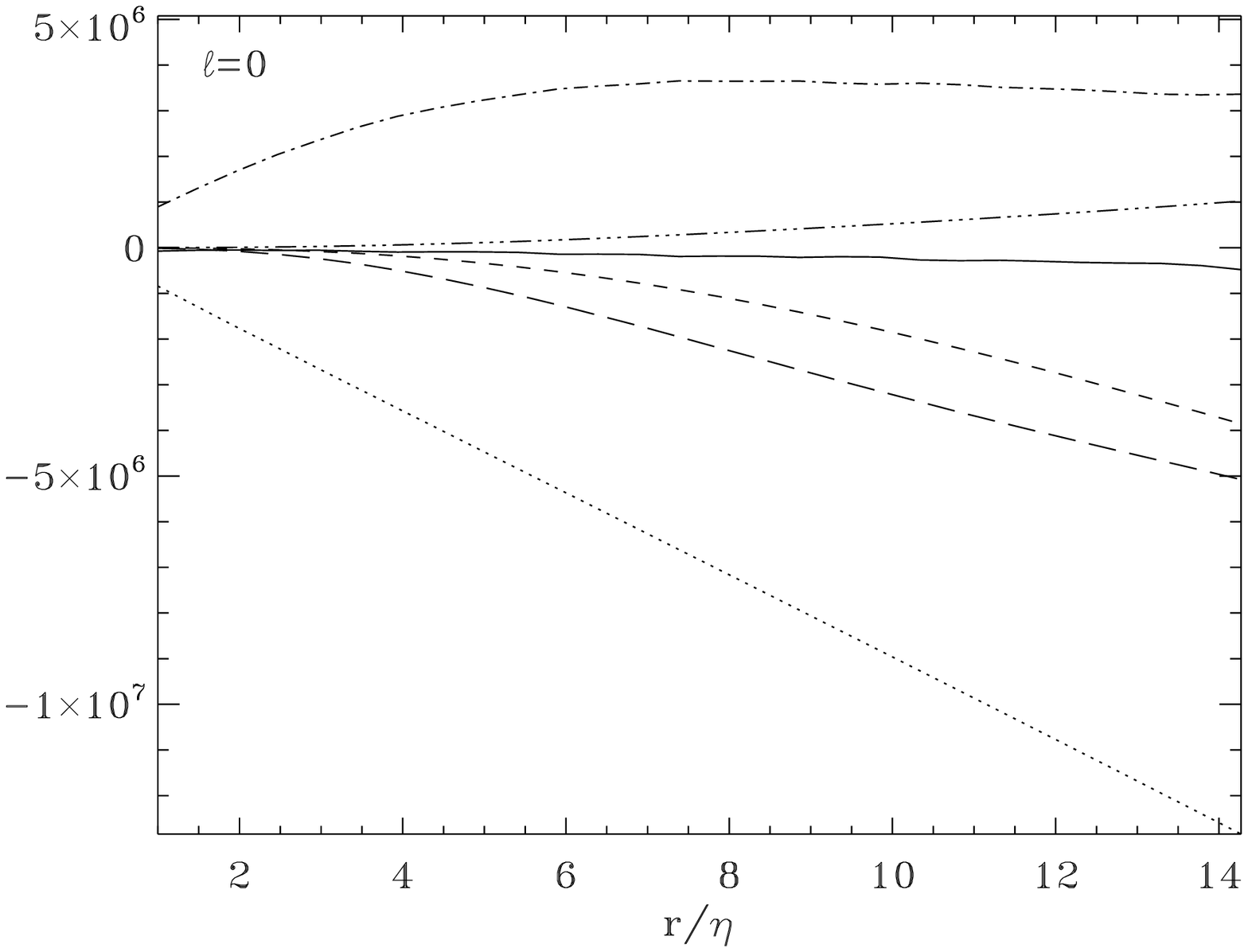}\hspace{-0.7cm}
\includegraphics[width=7.3cm]{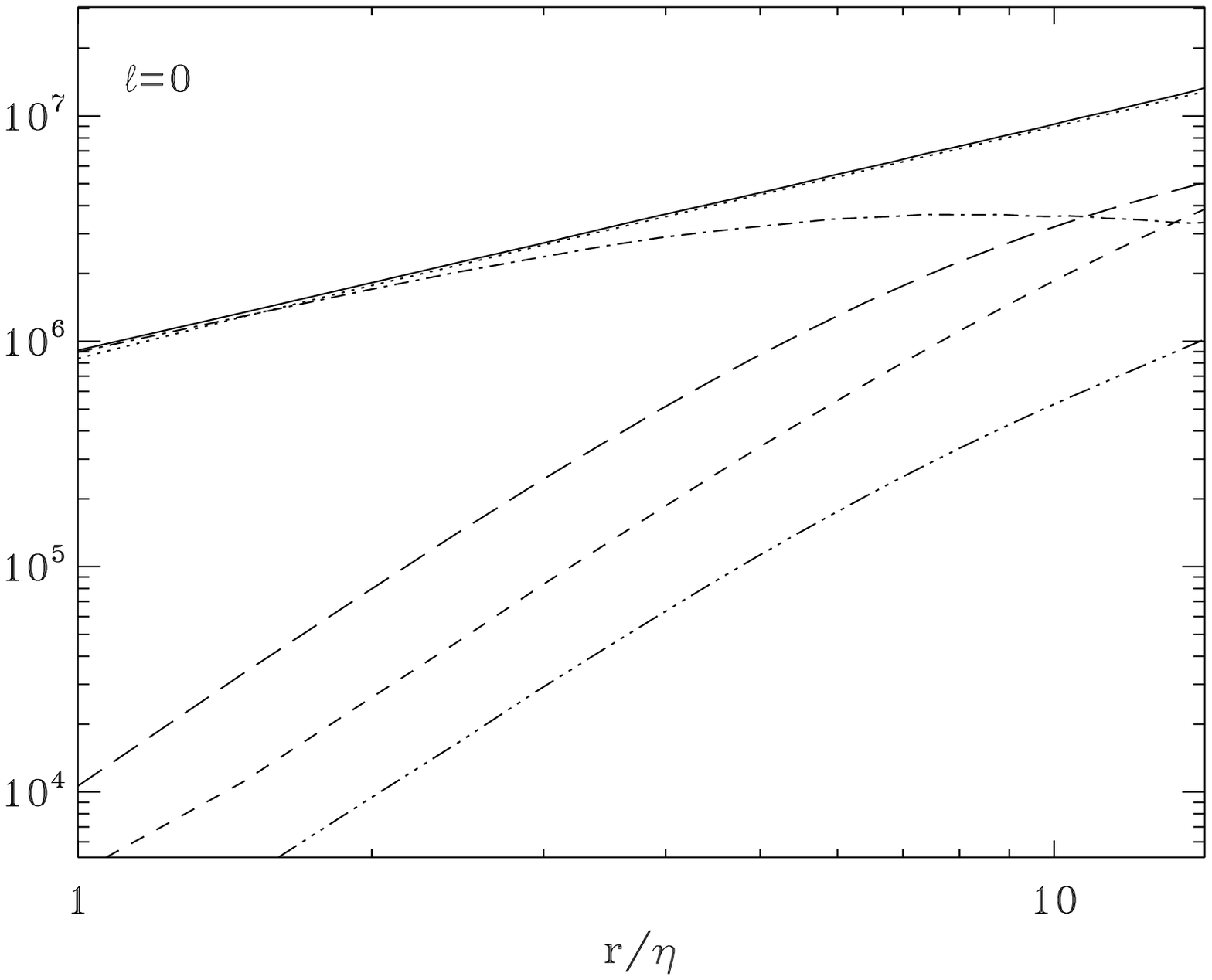}\hspace{-0.8cm}
  \caption{Two representations of equation~(\ref{eqVfinal_l=0}) for
    $z=1/4$. The legends are the same as in figure~\ref{figV_l=0}.}
  \label{figV_l=0off}
\end{figure}

\begin{figure}
\centering
\hspace{-1.cm}
\includegraphics[width=7.3cm]{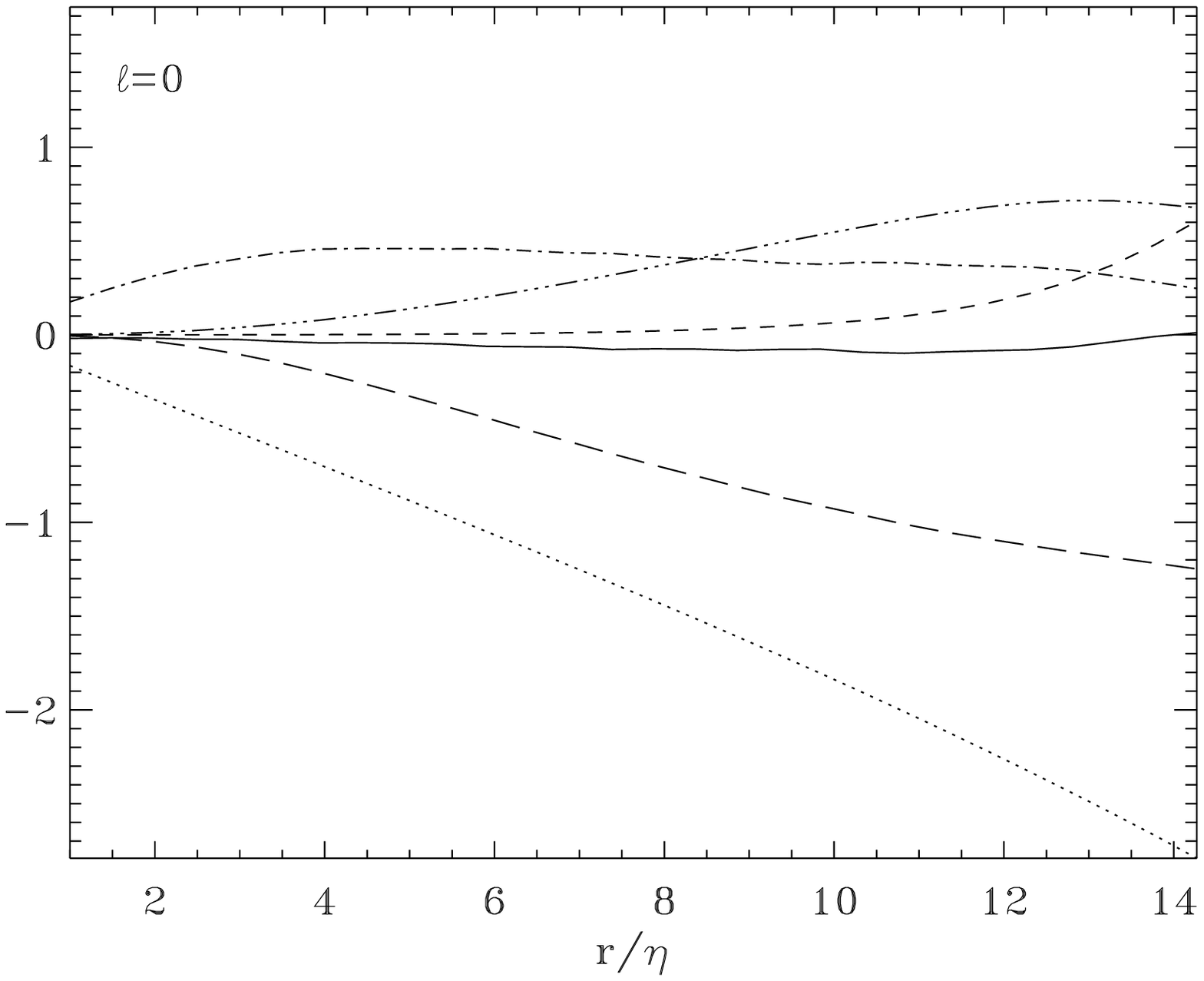}\hspace{-0.7cm}
\includegraphics[width=7.3cm]{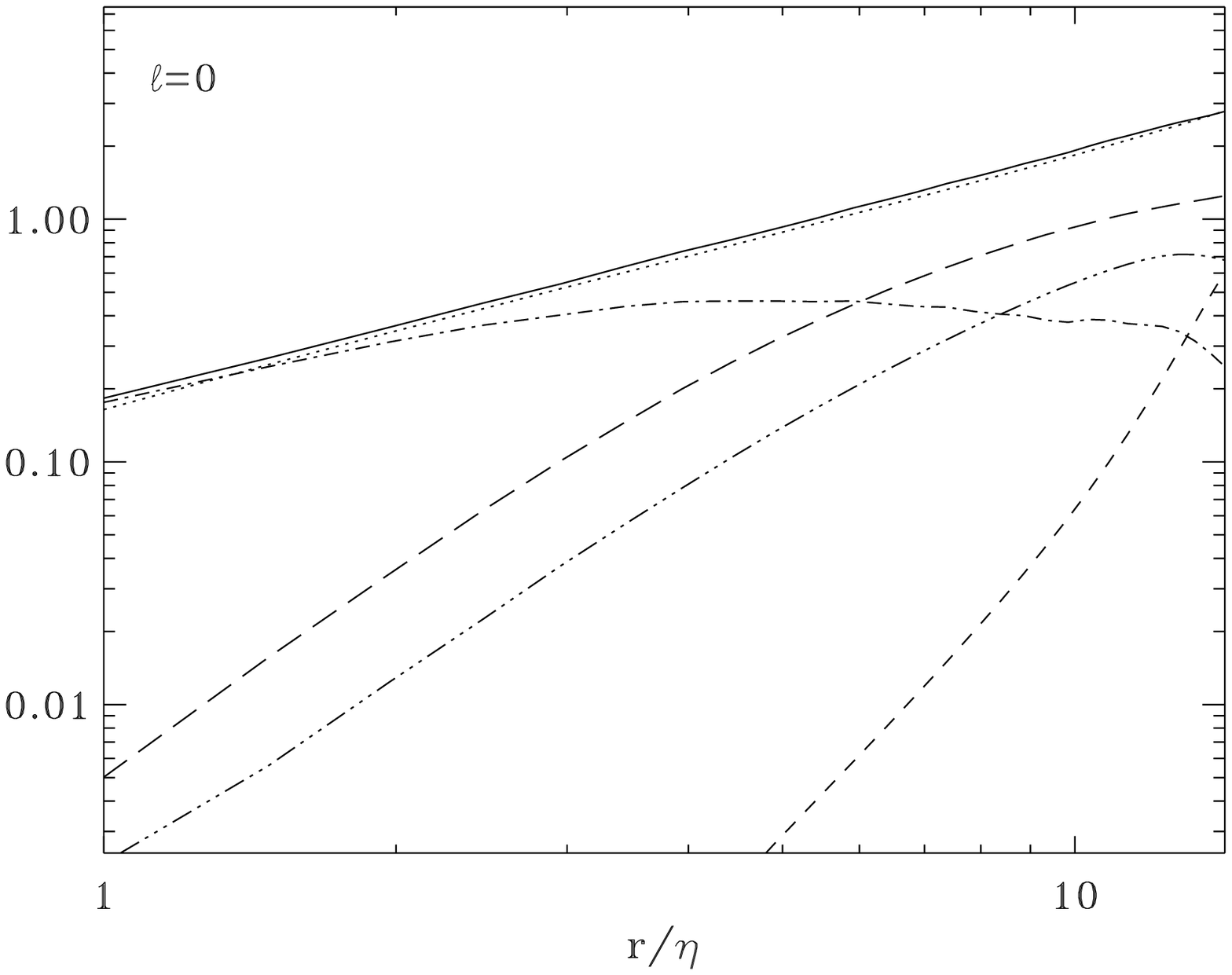}\hspace{-0.8cm}
  \caption{Two representations of 
   equation~(\ref{eqTfinal_l=0})  for
   $z=1/4$. The legends are the same as in figure~\ref{figT_l=0}.}
  \label{figT_l=0off}
\end{figure}

\begin{figure}
  \centering
 \includegraphics[height=7cm]{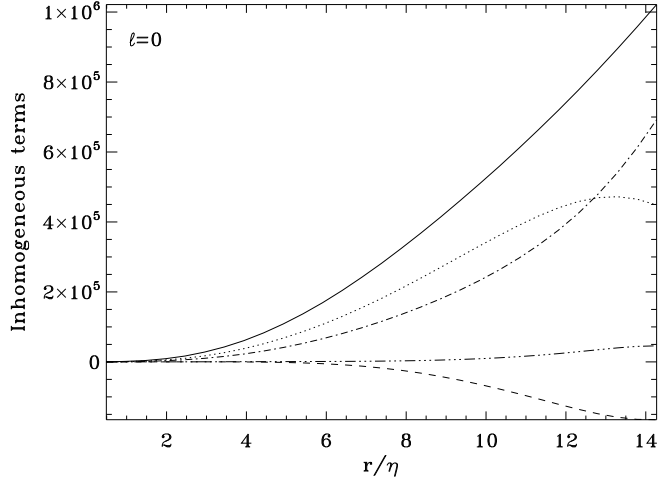}
  \caption{Decomposition of $\moy{NH}^0_0$ at $z=1/4$. Note the significant
    contribution of the pressure diffusion term. Same legends as in figure~\ref{decNHl=0}.}
  \label{decNHl=0off}
\end{figure}

\begin{figure}
  \centering
  \includegraphics[height=7cm]{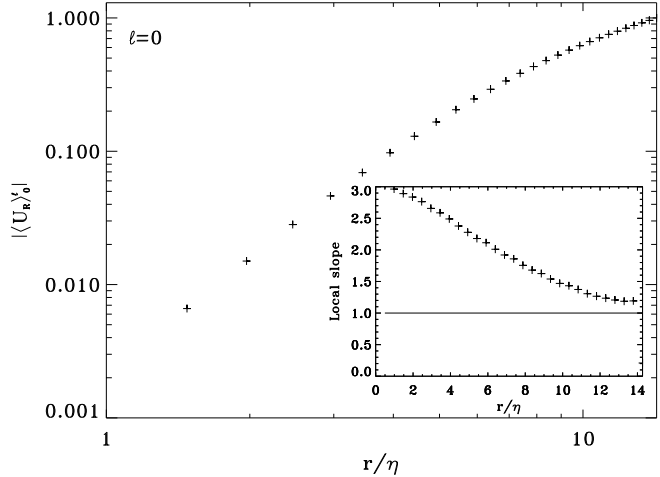}
  \caption{$|\moy{U_R}^0_0|$ normalized with respect to its maximum
    over $r$, as a function of $r/\eta$, at $z=1/4$. Inset:
    logarithmic slope of the same function.}
  \label{figS3l=0off}
\end{figure}

As can be seen on figure~\ref{figV_l=0off} and
figure~\ref{figT_l=0off}, the behaviour of the various terms in
equations~(\ref{eqVfinal_l=0})-(\ref{eqTfinal_l=0})
for $r<15 \eta$ and $z=1/4$ is not very different from the case
$z=1/2$ at first sight. 
There is however a slight difference
concerning inhomogeneous effects in the velocity equation, which  may
be due to the closeness of the bottom wall: figure~\ref{decNHl=0off}
reveals that normal pressure diffusion, which was small at small
scales for $z=1/2$, gives a significant contribution at the same
scales for $z=1/4$.  This term is notably
seen to continue to increase on the largest scales
available, while $\moy{NH_{u_z}}^0_0$ seems to have reached a maximum
for $r$ close to the distance to the bottom boundary $z$.
As far as the scaling of $\moy{U_R}^0_0$ is concerned,
figure~\ref{figS3l=0off} shows a possible trend towards slope 1
for $r/\eta\simeq 14$, but no real balance between the
$-4/3\moy{\varepsilon}r$ term and the structure function is visible on
figure~\ref{figV_l=0off}.

\begin{figure}
\centering
  \includegraphics[height=7.cm]{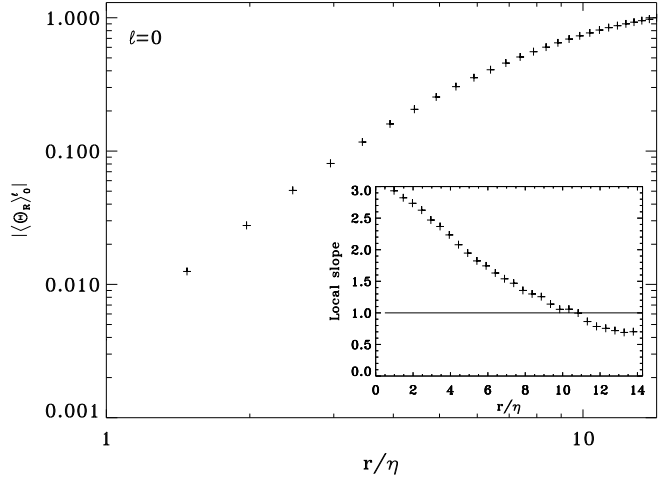}
  \caption{$|\moy{\Theta_R}^0_0|$ normalized with respect to its
    maximum over $r$ and its local logarithmic slope (inset),
    as a function of $r/\eta$, at $z=1/4$.}
\label{slopeThetaoff}
\end{figure}

\begin{figure}
  \centering
 \includegraphics[height=7cm]{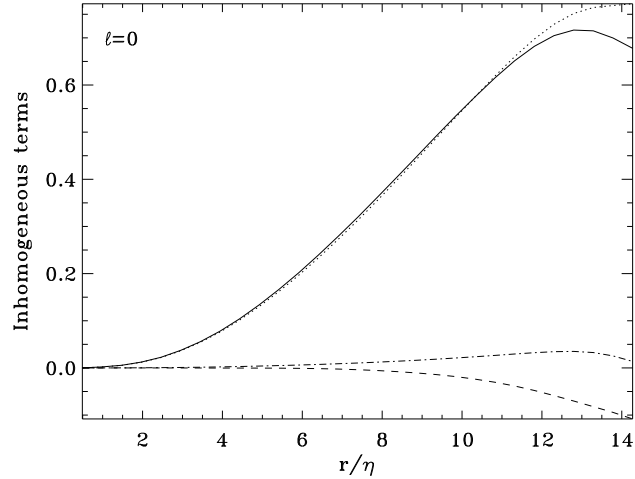}
  \caption{Decomposition of $\moy{NH^\theta}^0_0$ at $z=1/4$. Same
 legends as in figure~\ref{decNHTl=0}.}
  \label{decNHTl=0off}
\end{figure}

Meanwhile, figure~\ref{slopeThetaoff} reveals that the logarithmic slope of
$\moy{\Theta_R}^0_0$, unlike for $z=1/2$, does not seem
to tend to 1 at the largest scales available for
$z=1/4$. Figure~\ref{decNHTl=0off} shows that inhomogeneous effects in
the temperature equation are still mainly due to the
$\moy{NH^\theta_{u_z}}$ term in this region.
It therefore seems difficult to make definite conclusions regarding
scaling laws in this region. It can only be asserted, following
figure~\ref{figV_l=0off}, that the assumptions of homogeneity and of
large-scale forcing are still clearly violated at
$z=1/4$. Explaining the logarithmic slope 1 for $\moy{U_R}^0_0$ by
simple K41 arguments therefore proves difficult in that case. This
special slope may for instance be related to the penetration of
$\vec{r}$ in the lower boundary layers where the flow dynamics are
quite different from those in the bulk of the convection cell.

\subsection{Anisotropic sectors}
In the preceding paragraphs, two important assumptions of the K41
theory have been tested in the case of turbulent Rayleigh-B\'enard
convection. The homogeneity hypothesis has been shown to be
violated for $r/\eta>20$. The assumption of large-scale forcing has
been shown to be very crude, since buoyancy forcing remains important at
all non-dissipative scales of the simulation, even though the
effective Bolgiano length is comparable to
the depth of the convective layer. It has thus
been argued that the absence of scaling behaviour in turbulent
convection at $\ray=10^6$ was mostly related to these effects, and
that in the ideal situation of a homogeneous flow at
$\rey_\lambda=30$ forced on large scales only, inertial range scalings
should be observed. There is however a possible supplementary 
cause for the apparent lack of inertial range scalings in turbulent flows,
namely anisotropy. This has been demonstrated by
\cite{arad99b} for the channel flow and will be demonstrated
in this paragraph for Rayleigh-B\'enard convection. An important remark
is that in mildly turbulent numerical simulations, anisotropy may be
caused by different effects. 
It may  be due  to physical effects such as spherical symmetry
breaking by gravity or the presence of horizontal plates in the system,
as outlined in the introduction, but may also be related to the fact that
simulations are done in cartesian boxes that break the rotational
symmetry too.
 This phenomenon has been notably pointed out  by \cite*{biferale00}, who
 have shown that the projections of structure functions on anisotropic sectors were
non-vanishing for a DNS of homogeneous isotropic turbulence performed in
a cartesian domain. These various effects are
difficult to disentangle. In order to quantify the exact contribution of spurious
numerical anisotropy, it would be necessary to perform statistical
analyses at much higher resolutions which,  as mentioned in the section
about numerics, is unfortunately not  feasible currently. In every
numerical study of anisotropic effects in turbulence, 
one should therefore keep in mind that spurious effects are present
and consider the results with some caution.

\begin{figure}
\centering
\hspace{-0.85cm}
\includegraphics[width=7.3cm]{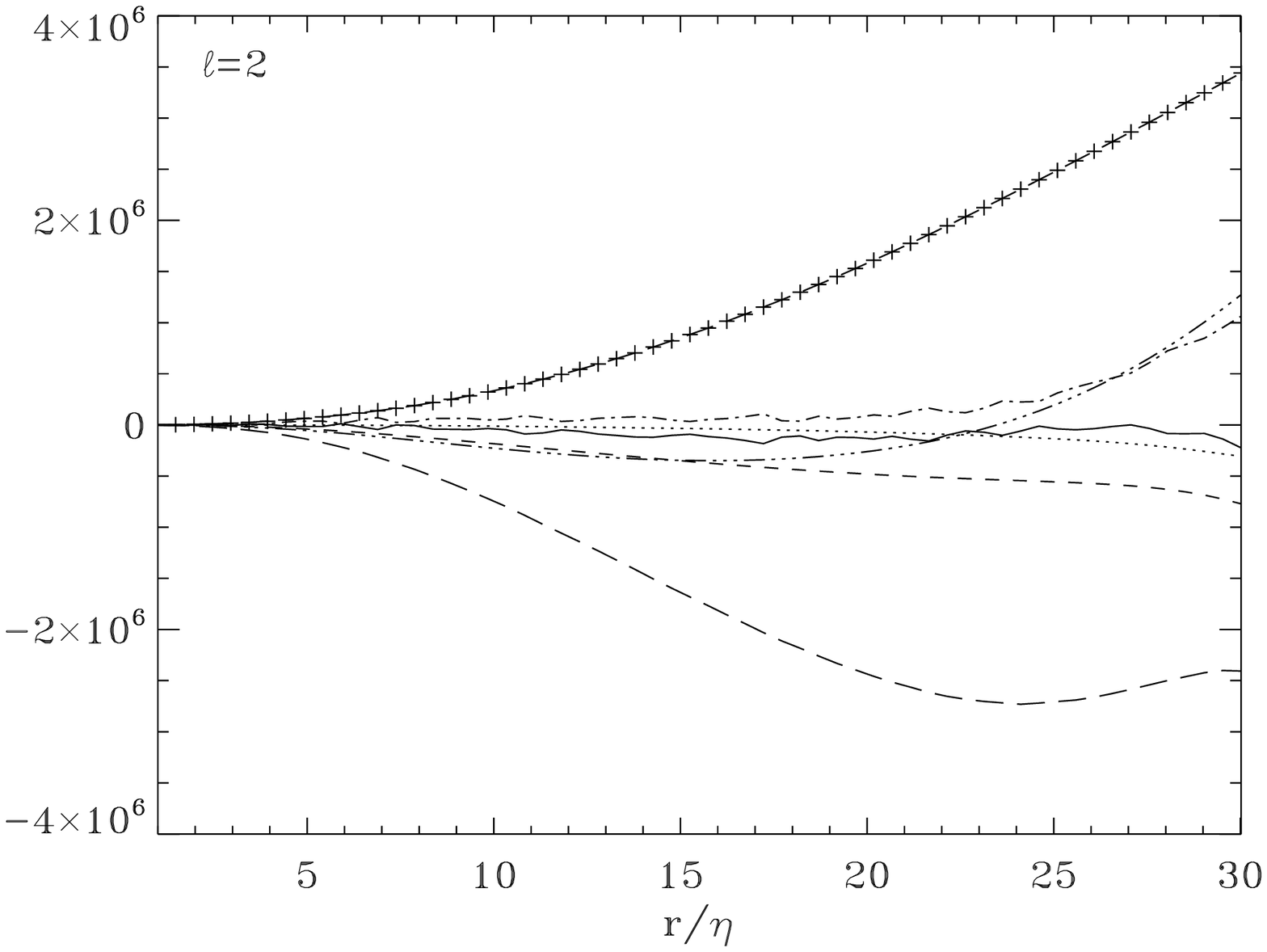}\hspace{-0.7cm}
\includegraphics[width=7.3cm]{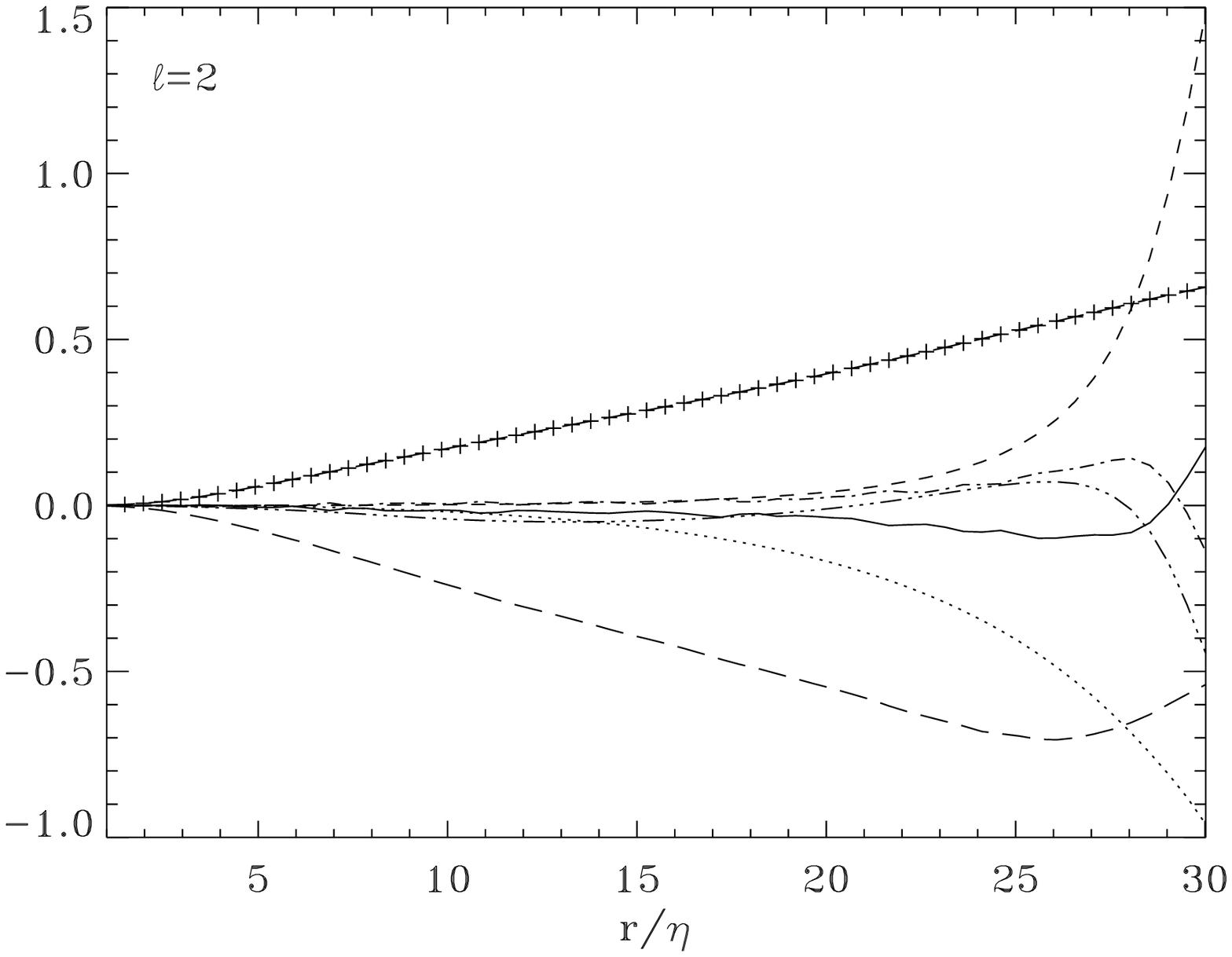}\hspace{-0.8cm}
  \caption{Left: equation~(\ref{eqVfinal}) for $\ell=2$ with
    $\moy{U_R}^\ell_0$ (\longdashline), $-\ell(\ell+1)/r^2\int_0^r
    y\moy{U_S}^{\ell}_0\mbox{d}y$ (+++++),
    $-2/r^2\int_0^r
    y^2\left[\moy{\varepsilon}+\moy{\varepsilon'}\right]^\ell_0\mbox{d}y$
    (\dotline), buoyancy term (\dashline), viscous term (\dashdotline), 
    $\moy{NH}^\ell_0$ (\dashtripledotline) and net budget (\fullline).
    Right: equation~(\ref{eqTfinal}) for $\ell=2$ with 
$\moy{\Theta_R}^\ell_0$ (\longdashline),  $-\ell(\ell+1)/r^2\int_0^r
y\moy{\Theta_S}^{\ell}_0\mbox{d}y$ (+++++), $-2/r^2\int_0^r
\!\! y^2\left[\moy{N}+\moy{N'}\right]^\ell_0\mbox{d}y$ (\dotline),
mean temperature advection term (\dashline), diffusion term (\dashdotline),
$\moy{NH^\theta}^\ell_0$ (\dashtripledotline) and net budget (\fullline).}
\label{figS3lneq0}
\end{figure}

The analysis is restricted to the central plane $z=1/2$.
The largest sectorial order $\ell$ considered in this paragraph is 6, for
statistical convergence reasons. The scale-by-scale budgets for
the anisotropic sectors read
 \begin{align}
\displaystyle{-\moy{U_R}^\ell_0+\f{\ell(\ell+1)}{r}\moy{U_S}^{\ell}_0+\f{2\alpha g}{r^2}\int_0^r
y^2\moy{ \Delta u_z\Delta \theta}^\ell_0\mbox{d}y} \qquad\qquad\qquad\qquad\qquad\qquad\notag  \\ \displaystyle{+2\nu\gradr{}{}\moy{(\Delta
    u_i)^2}^\ell_0-\f{2\nu\,\ell(\ell+1)}{r^2}\int_0^r\moy{(\Delta
  u_i)^2}^{\ell}_0\mbox{d}y+\moy{NH}^\ell_0} =\qquad\qquad\notag\\  \f{2}{r^2}\int_0^r
y^2\left[\moy{\varepsilon}+\moy{\varepsilon'}\right]^\ell_0\mbox{d}y
\label{eqVfinal_lneq0}
 \end{align}
and
 \begin{align}
   \label{eqTfinal_lneq0}
\displaystyle{-\moy{\Theta_R}^\ell_0+\f{\ell(\ell+1)}{r}\moy{\Theta_S}^{\ell}_0}
+\displaystyle{\f{2}{r^2}\int_0^r
  y^2\left(\moy{u_z\diffz{\moy{T}}\Delta\theta}^\ell_0- 
 \moy{u_z'\diffzp{\moy{T'}}\Delta\theta}^\ell_0\right)\mbox{d}y}
\qquad\qquad\notag 
\\ \displaystyle{+2\kappa\gradr{}{}\moy{(\Delta
    \theta)^2}^\ell_0-\f{2\kappa\,\ell(\ell+1)}{r^2}\int_0^r\moy{(\Delta
  \theta)^2}^\ell_0\mbox{d}y+\moy{NH\,^\theta}^\ell_0}  =
\qquad\qquad\qquad \notag\\
 \displaystyle{\f{2}{r^2}\int_0^r y^2\left[\moy{N}+\moy{N'}\right]^\ell_0\mbox{d}y}~.
 \end{align}
These budgets involve extra terms related to anisotropy in comparison
to equations~(\ref{eqVfinal_l=0})-(\ref{eqTfinal_l=0}). The most
important ones are those containing the transverse third-order
structure functions $\moy{U_S}^\ell_0$ and  $\moy{\Theta_S}^\ell_0$.
To illustrate this, the $\ell=2$ component of equations~(\ref{eqVfinal})-(\ref{eqTfinal})
can be investigated in more detail (figure~\ref{figS3lneq0}).
The (positive) transverse third-order structure function terms are shown 
to compensate significantly the (negative) longitudinal
third-order structure functions in each equation.
There is also a small anisotropic contribution of the
$\left[\moy{\varepsilon}+\moy{\varepsilon'}\right]^\ell_0$ and
$\left[\moy{N}+\moy{N'}\right]^\ell_0$ terms at the largest
$r$, which is due to increased dissipation in the boundary
layers. Finally, as was the case for the isotropic equations, the
inhomogeneous terms contribute to the scale-by-scale
budgets in the anisotropic sectors. 

\begin{figure}
  \centering
\includegraphics[height=7cm]{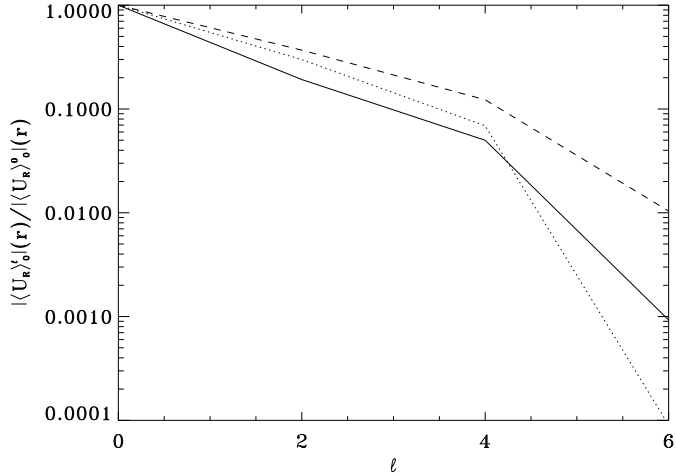}
  \caption{Spherical harmonics spectrum of $\moy{U_R}$ at the center
    of the convection cell for $r/\eta=7.9$ (\fullline), $r/\eta=15.3$
    (\dotline) and $r/\eta=22.6$ (\dashline). Spectra are normalized
    with respect to $\moy{U_R}^0_0(r)$ for each $r$. The $\ell=2$ and
    $\ell=4$ terms  are shown to  have a comparable magnitude to the
    $\ell=0$ one at moderate to large correlation lengths.}
  \label{spectrum_l}
\end{figure}

\begin{figure}
  \centering
\hbox{\hspace{-.45cm}\includegraphics[height=5.5cm]{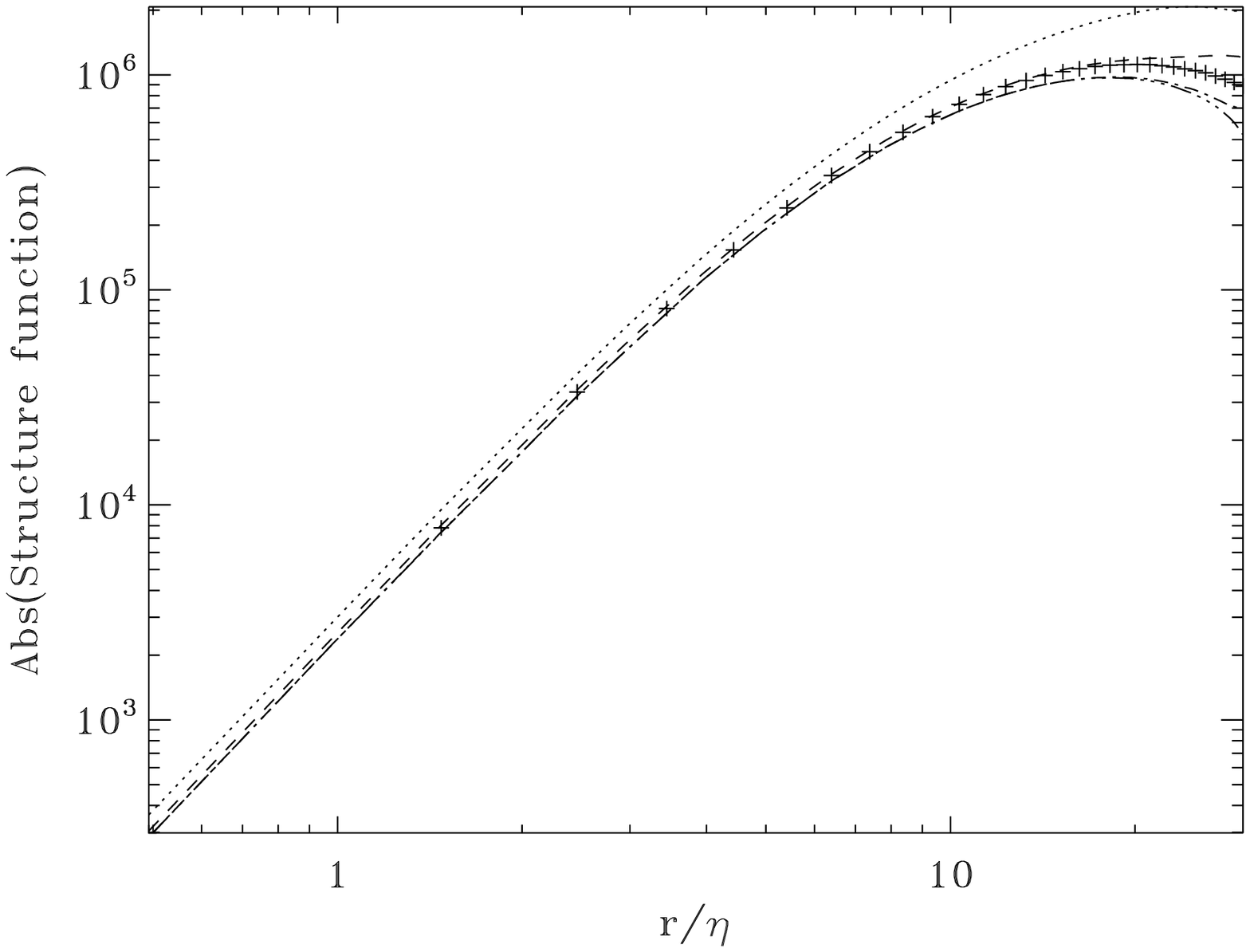} \hspace{-0.6cm}
\includegraphics[height=5.5cm]{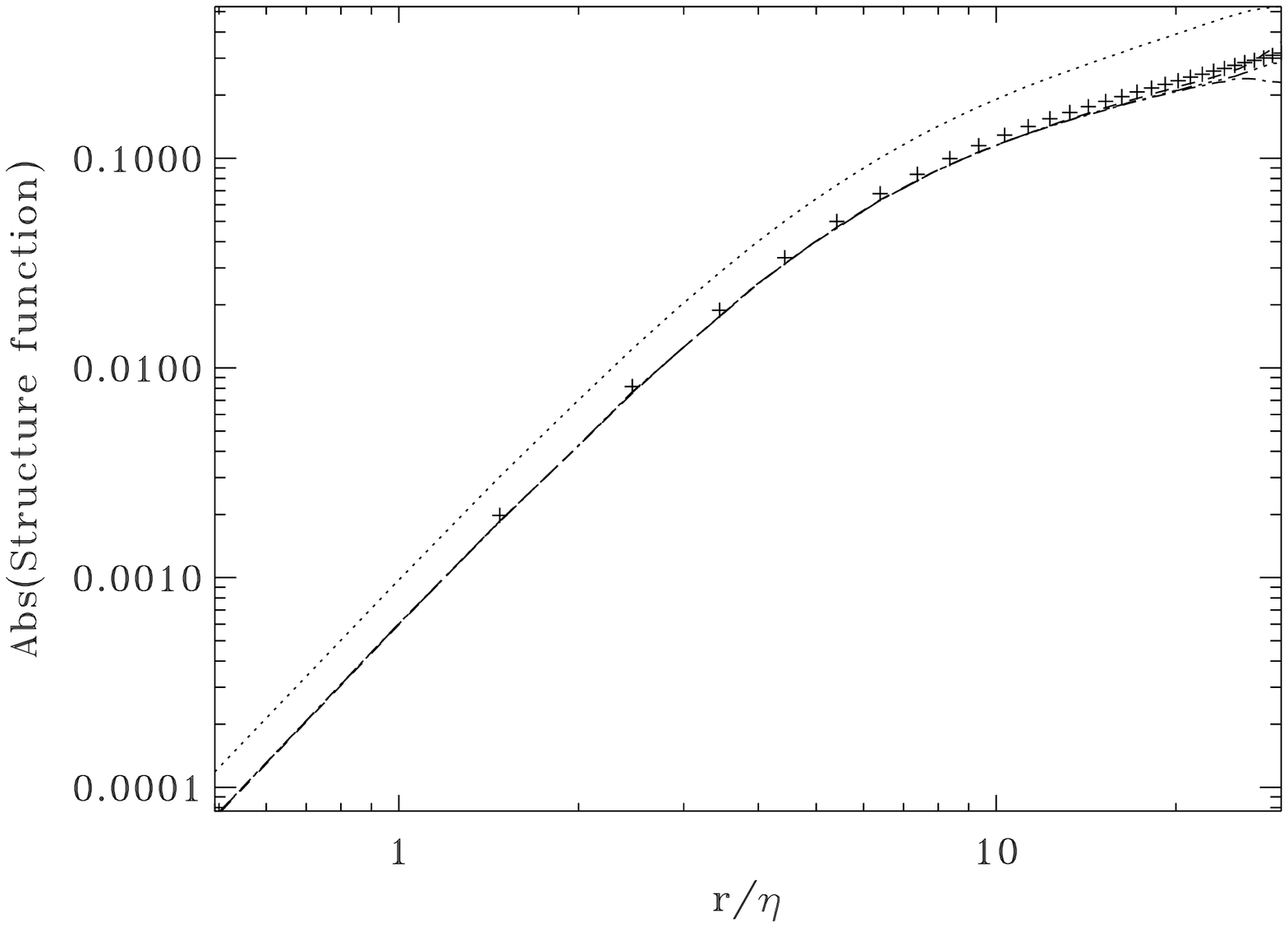}}
 \caption{Reconstruction of the reduced structure functions
    $\moy{U_r}(\theta=\pi/2)$ (left plot, + symbols) and
    $\moy{\Theta_r}(\theta=\pi/2)$ (right plot, + symbols) at the center
    of the convection cell using anisotropic components up to
    $\ell=0$ (\dotline), $\ell=2$ (\dashline), $\ell=4$ (\dashdotline),
    $\ell=6$ (\dashtripledotline). Note the significant contribution of the
    $\ell=2$ components to the reduced structure functions.}
  \label{reconstruct}
\end{figure}

\begin{figure}
  \centering
\hbox{\hspace{-.45cm}\includegraphics[height=5.5cm]{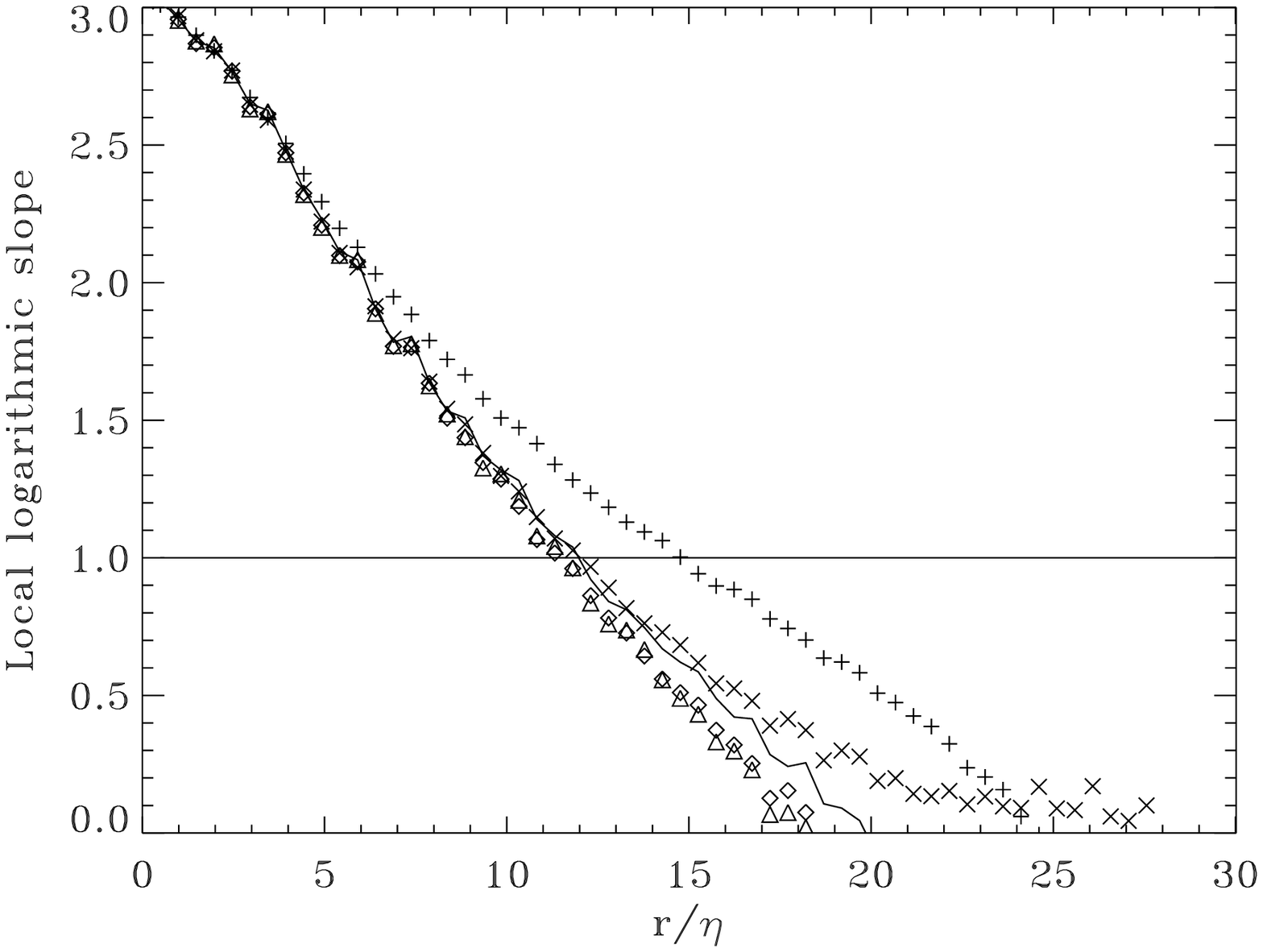} \hspace{-0.6cm}
 \includegraphics[height=5.5cm]{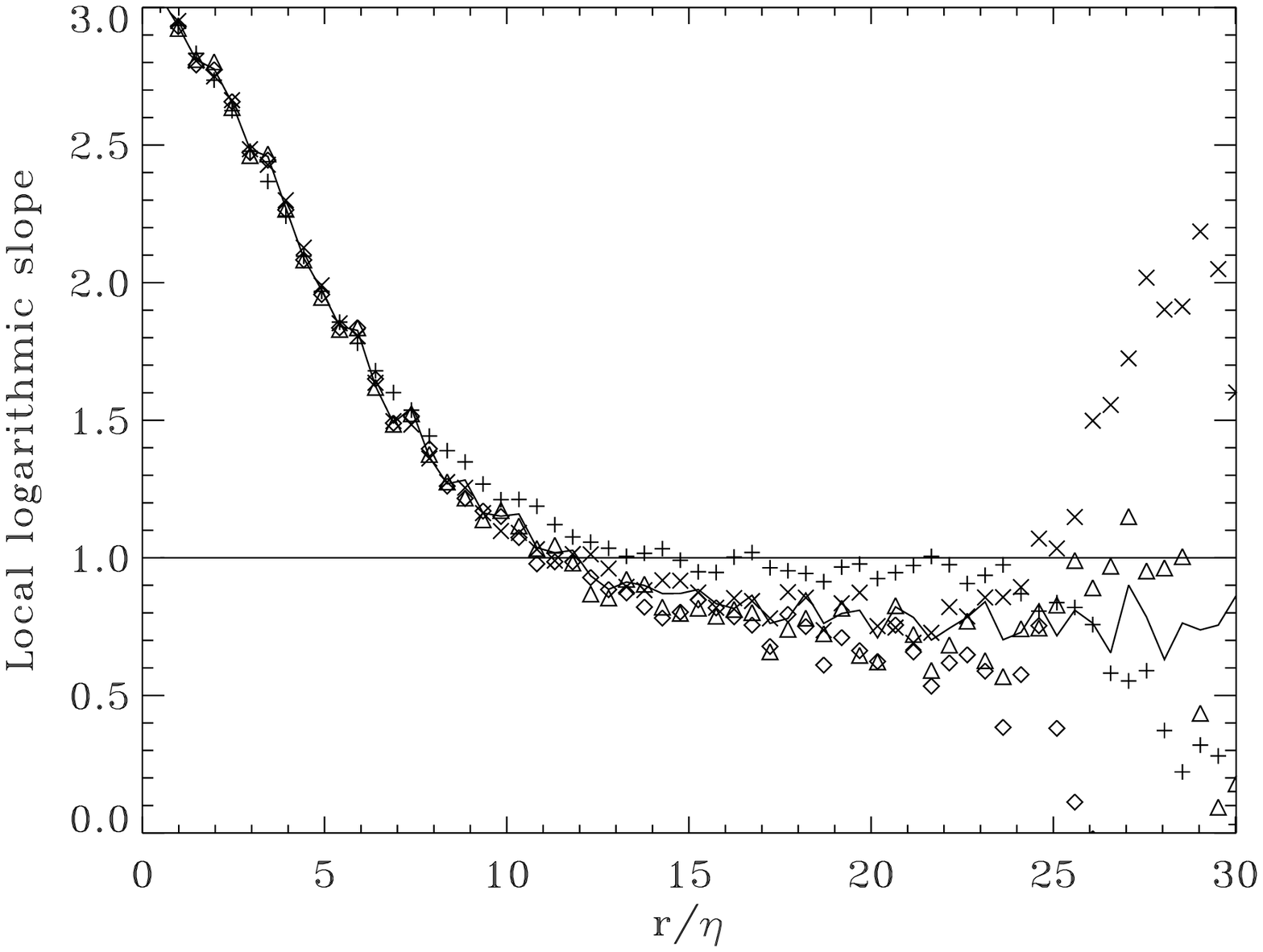}}
  \caption{Reconstruction of the local logarithmic slope of reduced
    structure functions $\moy{U_r}(\theta=\pi/2)$ (left plot,\fullline) 
    and $\moy{\Theta_r}(\theta=\pi/2)$ (right plot,\fullline) at the center
    of the convection cell, using anisotropic components up to
    $\ell=0$ (+), $\ell=2$ ($\times$), $\ell=4$ ($\diamond$), $\ell=6$
    ($\triangle$).}
  \label{reconstructslope}
\end{figure}

Figure~\ref{spectrum_l} displays the spherical harmonics spectrum
of $\moy{U_R}^\ell_0$ for even $\ell$ up to $\ell=6$ (remember that
odd $\ell$ contributions are zero  when considering averages in the
central plane). It is once again emphasized that this spectrum contains
contributions of both physical anisotropies due to gravity and spurious
numerical anisotropies due to the use of a cartesian computational domain.
The $\ell\neq 0$ components appear to be
significant in comparison to their isotropic counterpart at moderate
to large correlation lengths. As the complete third-order structure
functions result from a linear combination (with weights
given by the spherical harmonics functions) of these various
significant contributions, which may follow distinct scaling laws if
foliation occurs, it is expected that genuine inertial range scalings
predicted by isotropic theories should be difficult to observe for
reduced structure functions, which correspond to the
complete structure functions  taken at a polar angle
$\theta=\pi/2$ only and therefore involve the contribution of all even
spherical harmonics components of these complete structure functions
evaluated at this angle. This is confirmed by figure~\ref{reconstruct} and
figure~\ref{reconstructslope}: figure~\ref{reconstruct} shows a
reconstruction of the reduced structure functions
$\moy{U_r}(\theta=\pi/2)$ and $\moy{\Theta_r}(\theta=\pi/2)$ averaged
over $\varphi$ using the $\ell=0,2,4,6$ components of the complete
$\moy{U_r}=\moy{(\Delta u_i)^2\Delta u_r}$ and
$\moy{\Theta_r}=\moy{(\Delta \theta)^2\Delta u_r}$ for the present
simulation. Figure~\ref{reconstructslope} is a similar reconstruction of
the local logarithmic slopes of the same objects. The shape and local
logarithmic slopes of the reduced structure functions differ
substantially from those of the $\ell=0$  components, which shows that
anisotropic components (especially $\ell=2$, as can be seen on both
plots of figure~\ref{reconstruct}) have a marked influence on the
local scaling exponents of reduced structure functions. In the case
of $\moy{\Theta_r}$, one finds for instance that the local exponent
of the reduced structure function for $r/\eta=20$ is approximately 0.8,
while the $\ell=0$ component exponent is 1 at the same correlation
length,  as shown earlier in the paper and on the right plot of
figure~\ref{reconstructslope}. A similar analysis stands for $\moy{U_r}$ (left plot of
figure~\ref{reconstructslope}). Such differences can not be
ignored, since they are larger than typical intermittency corrections
to inertial range scalings and than differences between concurrent
theories of turbulence. Inertial range scaling exponents obtained via
an analysis of reduced structure functions should therefore be
regarded with caution, at least when the flow is fully
three-dimensional. For the same reasons, computations of structure functions in
laboratory experiments, which rely on single points measurements and
on the Taylor hypothesis along a specific  direction (that of the mean wind in
Rayleigh-B\'enard convection) will not necessarily lead to correct
measurements of scaling exponents predicted by isotropic theories or to
reliable values of intermittency corrections, because such exponents 
only apply to the isotropic component of the structure functions.
Note however that the previous conclusions do not mean that measuring
structure functions along a specific direction does not make
physical sense. For instance, using reduced structure functions
may prove useful to analyse scalings in two-dimensional flows, layered flows
or in strongly anisotropic flows such as boundary layers, where
scalings in the directions perpendicular to the shear can be
observed. Also, in regions where the SO(3) decomposition can not be
performed,  as is the case close to walls, the only currently available methods of
analysis are the computation of reduced structure functions and single
points measurements coupled to the Taylor hypothesis.


The present analysis finally proves useful to test theoretical arguments on
anisotropic turbulence and the phenomenon of foliation: according to
modern theories of anisotropy
\citep{arad99a,biferalephysrep}, each $\ell$ component of structure
functions may follow a distinct inertial range scaling law with a
scaling exponent increasing with $\ell$. Although it proves difficult to
identify scaling ranges (and therefore scaling exponents) on the
$\ell\neq 0$ sectors of the third-order structure functions in this
simulation, the increase of scaling exponents with $\ell$ seems to be qualitatively
well verified, at least it is consistent with the fact that
the $\ell=2$ contribution is more important for large $r$ than for small
$r$ in comparison to the $\ell=0$ component. This may be
viewed as an argument in favour of a small-scale return to
isotropy. Note however that definite conclusions can not be drawn
here because the $\ell=2$ component still remains significant
for small $r$ (figure~\ref{reconstruct}). This may be a finite
resolution effect related to the use of cartesian grids and boxes that
do not have rotational symmetry.

\section{\label{discuss}Summary and discussion}
New results on inertial range scaling laws, scale-by-scale budgets and
correlation functions in turbulent Rayleigh-B\'enard convection at moderate Rayleigh
and Reynolds numbers have been reported in this paper. A derivation of  generalized Kolmogorov
and Yaglom equations~(\ref{eqVfinal})-(\ref{eqTfinal}),
which notably makes use of the SO(3) decomposition of statistical
averages, has first been presented. The formalism provides a convenient way
of disentangling inhomogeneous and anisotropic effects in 
convective turbulence, but it can also be used 
to study other anisotropic and/or inhomogeneous
numerically simulated or experimental flows. For this purpose, the
buoyancy forcing term in the velocity equation should be replaced by 
any appropriate anisotropic forcing mechanism.

The analysis of scale-by-scale budgets in a  convection DNS at $\ray=10^6$
has led to the following conclusions. First of all, a compensation
effect between the classical
$-4/3\moy{\varepsilon}r$ term of the generalized Kolmogorov equation on one hand, inhomogeneous
terms, viscous terms and buoyancy forcing on the other hand, has been
demonstrated in the isotropic sector of the velocity equation. 
Among these three terms, buoyancy is the most important one. This
compensation effect prevents any clear scaling law from showing up
at $\rey_\lambda=30$ in the case of the third-order velocity structure function. 
Also, equation~({\ref{bolglength}) has been
  shown to underestimate significantly the
effective value of the Bolgiano length, at least for convective flows
in moderate aspect ratio containers.  
The analysis of the mixed temperature structure function
$\moy{\Theta_R}^0_0$ has revealed a scaling behaviour compatible with both K41 and BO59
theories in the central plane, in spite of the presence of
inhomogeneous effects on large scales. Meanwhile,
neglecting the term related to advection of the mean temperature profile, 
which is required in the BO59 theory, has been shown to be
justified. Scale-by-scale budgets computed off the central plane have
led to similar conclusions.
An interesting inhomogeneous effect at $z=1/4$ is the significant contribution of
pressure diffusion normal to the bottom wall.

A spherical harmonics spectrum of structure functions has finally revealed 
that the low $\ell$ degree contributions to the third-order structure functions
are not small in comparison to their isotropic counterparts at
moderate to large scales. It has particularly been shown that using reduced
structure functions at fixed polar angle in the bulk of a convection
cell in order to reveal scaling behaviour predicted by isotropic theories is misleading,
since these structure functions involve linear combinations of various $\ell$
components which scale differently with respect to $r$. This argument also
indicates that using single points measurements  in the bulk of
three-dimensional flows  together with the Taylor hypothesis in the
particular direction of a mean flow to test the predictions of
asymptotic dimensional isotropic theories of turbulence
or to calculate intermittency corrections to  these theories may lead
to significant biases. It has however been outlined that using
reduced structure functions may still be appropriate when the direction along
which these functions are computed has a special importance, as is the
case for instance in layered flows or boundary layers. Also, it has been
pointed out that spurious anisotropies related to the use of cartesian domains in 
simulations of mildly turbulent flows, which are difficult to quantify,
certainly interfere with the genuine anisotropic effects due to gravity.
Thus, the respective amplitudes of the various $\ell$ components of structure
functions at  moderate Reynolds number has to be considered with some caution.
Independently of the problem of the origin of anisotropies, these results
confirm the analysis of \cite{arad99b}
for the channel flow: disentangling anisotropic effects can not be avoided
if one wants to study the scaling behaviour of 
anisotropic systems correctly. Besides, even though the study of the
$r$-dependence of different $\ell$ components in
equations~(\ref{eqVfinal})-(\ref{eqTfinal}) seems
to indicate qualitatively small-scale return to isotropy in the case
of Rayleigh-B\'enard convection, it can not be ruled out that
anisotropic effects  persist at small scales, even at very high
Reynolds numbers. Small-scale anisotropies (called ``ramp-cliff
structures'') have for instance been reported in studies of turbulent
mixing of passive scalars (\cite{mestayer76,antonia78}, see also
\cite{warhaft2000}).

An important issue is to which extent the present results can be used to 
infer scaling behaviour at very high Rayleigh and Reynolds numbers. 
The still important contribution  of buoyancy forcing to the scale-by-scale
budget at scales smaller than $L_B$ naturally raises the question
of how many scale decades are required in a turbulent convective flow
in order to be able to identify definite scalings. 
In the soft turbulence regime, where $\nuss\sim \ray^{\gamma}$ with
$1/4<\gamma<1/3$ \citep*{grossmann00}, the Bolgiano length should not
depend significantly on $\ray$. For instance, following equation~(\ref{bolglength}),
$L_B\sim\ray^{-3/28}$ is obtained for $\gamma=2/7$ \citep{procaccia89}. 
In the hard regime where it has recently been shown that $\nuss\sim
(\ray\pr)^{1/2}$ \citep{calzavarini05}, the Bolgiano length should
remain constant (with respect to both $\ray$ and $\pr$ in that case, while the $\pr$
dependence in the hard turbulence regime depends on whether $\pr>1$ or
$\pr<1$), thus leading to the same result at all Rayleigh numbers.
For the present aspect ratio and Prandtl number, $L_B$ should
therefore remain close to 1 at all Rayleigh numbers.
On the contrary, the ratio between the dissipation scale and the
effective Bolgiano scale is expected to increase with increasing
Rayleigh number, which should in principle help to observe K41 scalings
if $\eta\ll r\ll L_B$ can be achieved \citep*{grossmannlvov93}. One
should however derive more precise conditions of applicability of K41 in that case. 
Assuming K41 to be valid for $r\leq L_B$, the buoyancy term, which equals the
$-4/3\moy{\varepsilon} r$ term at $L_B$, would scale like $r^{5/3}$
in that range. This approximation looks rather crude for $r$ close to
$L_B$, but as the buoyancy term should scale like $r^{9/5}$  above 
$L_B$ according to BO59, the actual local scaling exponent close to
$L_B$ may not be very different
from this 5/3 value. The important point here is that this exponent be
larger than 1 (which is the exponent of the $-4/3\moy{\varepsilon} r$
term). According to these scalings, non-dissipative scales as small as
$3\times10^{-2} L_B$ should be available in order for the buoyancy
term to become ten times smaller than the $-4/3\moy{\varepsilon} r$
term, which roughly corresponds to the conditions of the present
simulation, in which not definite K41 plateau can be observed
(figure~\ref{figS3l=0}). In order to find a range of scales in which
the buoyancy term would become at least one hundred times
smaller than the $-4/3\moy{\varepsilon}r$ term, one should then
look for a regime where $\eta<10^{-3} L_B$. The required Rayleigh
number can be evaluated by using
$\eta\sim \ray^{-9/28}$, proposed by \cite{grossmannlohse93}. This relation
 can be calibrated using the results of the present simulation at
 $\ray=10^6$, for which $\eta=0.016$.
One finally finds that $\ray$ has to exceed several times $10^9$ to obtain a
flow with $\eta< 10^{-3} L_B$. Even in this situation, definite
scalings are expected only in the subrange $10\eta<r<0.1 L_B$,
that is over one decade. The main problem is that two scale
separations  $r\ll L_B$ and $r\gg\eta$ must simultaneously be
satisfied, which can only occur at very high Rayleigh number. This is
an important argument to understand why inertial range scalings have
been difficult to identify in experimental or numerical turbulent
convection studies.  

For the preceding reasons, it is probable that various
results on scaling exponents obtained in the range $10^6<\ray<10^9$
(\textit{e.~g.} \cite{calzavarini02}, who have computed local scaling
exponents from reduced structure functions) are biased simultaneously by 
anisotropic effects and significant buoyancy forcing at $r<L_B$.
The lack of complete K41 or BO59 scalings in the simulations at very
high Rayleigh number (up to $\ray=2\times 10^{11}$) of
\cite{verzicco03} may also be a hint that inertial range scalings in 
anisotropic flows such as Rayleigh-B\'enard convection can not be
understood easily using isotropic theories.
If inhomogeneity and isotropy were to play an important role in the
scaling behaviour of very high Rayleigh number convective turbulence,
the present results could serve as  guidelines to study them in detail.
More generally, the SO(3) decomposition
of structure functions appears to be a very powerful (and necessary)
tool to study various anisotropic flows.
It should be particularly helpful to use it
when possible in  research domains such as astrophysics or
atmospheric sciences, where  anisotropic turbulence is
ubiquitous. Such an investigation for strongly stratified
non-Boussinesq turbulent convection, which is relevant to the Sun and
other stars, is currently underway.

\smallskip

The author acknowledges several fruitful discussions with
A.~A.~Schekochihin and thanks F. Anselmet, F. Ligni\`eres and
M. Rieutord for many insightful comments and criticisms on early
versions of the manuscript. Numerical simulations have been performed
on the IBM SP4 supercomputer of  Institut du D\'eveloppement et des
Ressources en Informatique Scientifique (IDRIS, Orsay, France), which
is gratefully acknowledged. 

\medskip

\medskip

\medskip

\bibliographystyle{jfm}
\bibliography{rincon_bib}

\begin{thebibliography}{45}
\expandafter\ifx\csname natexlab\endcsname\relax\def\natexlab#1{#1}\fi

\bibitem[{Antonia} \& {Van Atta}(1978)]{antonia78}
{\sc {Antonia}, R. \& {Van Atta}, C.} 1978 Structure functions of temperature
  fluctuations in turbulent shear flows. {\em J. Fluid Mech.\/} {\bf 84},
  561--580.

\bibitem[Antonia {\em et~al.\/}(1997)Antonia, Ould-Rouis, Anselmet \&
  Zhu]{antonia97}
{\sc Antonia, R.~A., Ould-Rouis, M., Anselmet, F. \& Zhu, Y.} 1997 Analogy
  between predictions of {K}olmogorov and {Y}aglom. {\em J. Fluid Mech.\/} {\bf
  332}, 395--409.

\bibitem[{Arad} {\em et~al.\/}(1999{\natexlab{{\em a\/}}}){Arad}, {Biferale},
  {Mazzitelli} \& {Procaccia}]{arad99b}
{\sc {Arad}, I., {Biferale}, L., {Mazzitelli}, I. \& {Procaccia}, I.}
  1999{\natexlab{{\em a\/}}} {Disentangling scaling properties in anisotropic
  and inhomogeneous turbulence}. {\em Phys. Rev. Lett.\/} {\bf 82}, 5040--5043.

\bibitem[{Arad} {\em et~al.\/}(1999{\natexlab{{\em b\/}}}){Arad}, {L'vov} \&
  {Procaccia}]{arad99a}
{\sc {Arad}, I., {L'vov}, V. \& {Procaccia}, I.} 1999{\natexlab{{\em b\/}}}
  {Correlation functions in isotropic and anisotropic turbulence: the role of
  the symmetry group}. {\em Phys. Rev. E\/} {\bf 59}, 6753--6765.

\bibitem[{Benzi} {\em et~al.\/}(1993){Benzi}, {Ciliberto}, {Tripiccione},
  {Baudet}, {Massaioli} \& {Succi}]{benzi93}
{\sc {Benzi}, R., {Ciliberto}, S., {Tripiccione}, R., {Baudet}, C.,
  {Massaioli}, F. \& {Succi}, S.} 1993 {Extended self-similarity in turbulent
  flows}. {\em Phys. Rev. E\/} {\bf 48}, 29.

\bibitem[{Benzi} {\em et~al.\/}(1998){Benzi}, {Toschi} \&
  {Tripiccione}]{benzi98}
{\sc {Benzi}, R., {Toschi}, F. \& {Tripiccione}, R.} 1998 {On the heat transfer
  in Rayleigh-B\'enard systems}. {\em J. Stat. Phys.\/} {\bf 93}.

\bibitem[{Benzi} {\em et~al.\/}(1994){Benzi}, {Tripiccione}, {Massaioli},
  {Succi} \& {Ciliberto}]{benzi94}
{\sc {Benzi}, R., {Tripiccione}, R., {Massaioli}, F., {Succi}, S. \&
  {Ciliberto}, S.} 1994 On the scaling of the velocity and temperature
  structure functions in {R}ayleigh-{B}\'enard convection. {\em Europhys.
  Lett.\/} {\bf 25}~(5), 341--346.

\bibitem[{Biferale} {\em et~al.\/}(2003){Biferale}, {Calzavarini}, {Toschi} \&
  {Tripiccione}]{biferale03}
{\sc {Biferale}, L., {Calzavarini}, E., {Toschi}, F. \& {Tripiccione}, R.} 2003
  {Universality of anisotropic fluctuations from numerical simulations of
  turbulent flows}. {\em Europhys. Lett.\/} {\bf 64}, 461--467.

\bibitem[{Biferale} {\em et~al.\/}(2000){Biferale}, {Gualtieri} \&
  {Toschi}]{biferale00}
{\sc {Biferale}, L., {Gualtieri}, P. \& {Toschi}, F.} 2000 Statistics of
  pressure and of pressure-velocity correlations in isotropic turbulence. {\em
  Phys. Fluids\/} {\bf 12}~(7), 1836.

\bibitem[{Biferale} \& {Procaccia}(2005)]{biferalephysrep}
{\sc {Biferale}, L. \& {Procaccia}, I.} 2005 Anisotropy in turbulent flows and
  in turbulent transport. {\em Phys. Reports\/} {\bf 414}~(2), 43.

\bibitem[{Bolgiano}(1959)]{bolgiano59}
{\sc {Bolgiano}, R.} 1959 {Structure of turbulence in stratified media}. {\em
  J. Geophys. Res.\/} {\bf 64}, 2226.

\bibitem[{Califano}(1996)]{cali}
{\sc {Califano}, F.} 1996 {A numerical algorithm for geophysical and
  astrophysical inhomogeneous fluid flows}. {\em Comp. Phys. Comm.\/} {\bf 99},
  29.

\bibitem[{Calzavarini} {\em et~al.\/}(2005){Calzavarini}, {Lohse}, {Toschi} \&
  {Tripiccione}]{calzavarini05}
{\sc {Calzavarini}, E., {Lohse}, D., {Toschi}, F. \& {Tripiccione}, R.} 2005
  Rayleigh and {P}randtl number scaling in the bulk of {R}ayleigh-{B}\'enard
  turbulence. {\em Phys. Fluids\/} {\bf 17}~(5), 055107.

\bibitem[{Calzavarini} {\em et~al.\/}(2002){Calzavarini}, {Toschi} \&
  {Tripiccione}]{calzavarini02}
{\sc {Calzavarini}, E., {Toschi}, F. \& {Tripiccione}, R.} 2002 {Evidences of
  Bolgiano scaling in 3D Rayleigh-B\'enard convection}. {\em Phys. Rev. E\/}
  {\bf 66}, 016304.

\bibitem[Casciola {\em et~al.\/}(2003)Casciola, Gualtieri, Benzi \&
  Piva]{casciola03}
{\sc Casciola, C.~M., Gualtieri, P., Benzi, R. \& Piva, R.} 2003 Scale-by-scale
  budget and similarity laws for shear turbulence. {\em J. Fluid Mech.\/} {\bf
  476}, 105--114.

\bibitem[{Cattaneo} {\em et~al.\/}(1991){Cattaneo}, {Brummell}, {Toomre},
  {Malagoli} \& {Hurlburt}]{cattaneo91}
{\sc {Cattaneo}, F., {Brummell}, N.~H., {Toomre}, J., {Malagoli}, A. \&
  {Hurlburt}, N.~E.} 1991 {Turbulent compressible convection}. {\em Astrophys.
  J.\/} {\bf 370}, 282--294.

\bibitem[{Cattaneo} {\em et~al.\/}(2001){Cattaneo}, {Lenz} \&
  {Weiss}]{cattaneo01}
{\sc {Cattaneo}, F., {Lenz}, D. \& {Weiss}, N.} 2001 {On the origin of the
  solar mesogranulation}. {\em Astrophys. J.\/} {\bf 563}, L91--L94.

\bibitem[{C}handrasekhar(1961)]{chandra61}
{\sc {C}handrasekhar, S.} 1961 {\em {H}ydrodynamic and hydromagnetic
  stability\/}. {D}over.

\bibitem[{Chill{\' a}} {\em et~al.\/}(1993){Chill{\' a}}, {Ciliberto},
  {Innocenti} \& {Pampaloni}]{chilla93}
{\sc {Chill{\' a}}, F., {Ciliberto}, S., {Innocenti}, C. \& {Pampaloni}, E.}
  1993 {Boundary layer and scaling properties in turbulent thermal convection}.
  {\em Nuovo Cimento\/} {\bf 15D}~(9), 1229.

\bibitem[{Ching} {\em et~al.\/}(2004){Ching}, {Chui}, {Shang}, {Qiu}, {Tong} \&
  {Xia}]{ching04}
{\sc {Ching}, E. S.~C., {Chui}, K.~W., {Shang}, X.-D., {Qiu}, X.-L., {Tong}, P.
  \& {Xia}, K.-Q.} 2004 Velocity and temperature cross-scaling in turbulent
  thermal convection. {\em J. Turbulence\/} {\bf 5}.

\bibitem[Danaila {\em et~al.\/}(1999)Danaila, Anselmet, Zhou \&
  Antonia]{danaila99}
{\sc Danaila, L., Anselmet, F., Zhou, T. \& Antonia, R.~A.} 1999 A
  generalization of {Y}aglom's equation which accounts for the large-scale
  forcing in heated decaying turbulence. {\em 391\/} {\bf 391}, 359--372.

\bibitem[Danaila {\em et~al.\/}(2001)Danaila, Anselmet, Zhou \&
  Antonia]{danaila01}
{\sc Danaila, L., Anselmet, F., Zhou, T. \& Antonia, R.~A.} 2001 Turbulent
  energy scale budget equations in a fully developed channel flow. {\em J.
  Fluid Mech.\/} {\bf 430}, 87--109.

\bibitem[{Demuren} {\em et~al.\/}(2001){Demuren}, {Wilson} \&
  {Carpenter}]{demuren01}
{\sc {Demuren}, A.~O., {Wilson}, R.~V. \& {Carpenter}, M.} 2001 {Higher-order
  compact schemes for numerical simulation of incompressible flows, Part I:
  theoretical development}. {\em Numerical Heat Transfer\/} {\bf 39}, 207--230.

\bibitem[{Grossmann} \& {Lohse}(1993)]{grossmannlohse93}
{\sc {Grossmann}, S. \& {Lohse}, D.} 1993 {Characteristic scales in
  Rayleigh-B\'enard turbulence}. {\em Phys. Lett. A\/} {\bf 173}, 58.

\bibitem[{Grossmann} \& {Lohse}(2000)]{grossmann00}
{\sc {Grossmann}, S. \& {Lohse}, D.} 2000 Scaling in thermal convection: a
  unifying theory. {\em J. Fluid Mech.\/} {\bf 407}, 27.

\bibitem[{Grossmann} \& {L'vov}(1993)]{grossmannlvov93}
{\sc {Grossmann}, S. \& {L'vov}, V.} 1993 {Crossover of spectral scaling in
  thermal turbulence}. {\em Phys. Rev. E\/} {\bf 47}, 4161.

\bibitem[{Hartlep} {\em et~al.\/}(2003){Hartlep}, {Tilgner} \&
  {Busse}]{hartlep03}
{\sc {Hartlep}, T., {Tilgner}, A. \& {Busse}, F.~H.} 2003 {Large scale
  structures in {R}ayleigh-{B}{\' e}nard convection at high {R}ayleigh
  numbers}. {\em Phys. Rev. Lett.\/} {\bf 91}~(6), 064501.

\bibitem[{Heslot} {\em et~al.\/}(1987){Heslot}, {Castaing} \&
  {Libchaber}]{heslot87}
{\sc {Heslot}, F., {Castaing}, B. \& {Libchaber}, A.} 1987 {Transitions to
  turbulence in helium gas}. {\em Phys. Rev. A\/} {\bf 36}, 5870--5873.

\bibitem[Hill(1997)]{hill97}
{\sc Hill, R.~J.} 1997 Applicability of {K}olmogorov's and {M}onin's equations
  of turbulence. {\em J. Fluid Mech.\/} {\bf 353}, 67--81.

\bibitem[{Hill}(2002)]{hill2002}
{\sc {Hill}, R.~J.} 2002 {The approach of turbulence to the locally homogeneous
  asymptote as studied using exact structure-function equations}. {\em
  arXiv:physics/0206034\/} .

\bibitem[{Kolmogorov}(1941)]{k41}
{\sc {Kolmogorov}, A.~N.} 1941 Dissipation of energy in locally isotropic
  turbulence. {\em Dokl. Akad. Nauk SSSR\/} {\bf 32}, 16.

\bibitem[{Lele}(1992)]{lele92}
{\sc {Lele}, S.~K.} 1992 {Compact finite difference schemes with spectral-like
  resolution}. {\em J. Comp. Phys.\/} {\bf 103(1)}, 16--42.

\bibitem[Lindborg(1996)]{lindborg96}
{\sc Lindborg, E.} 1996 A note on {K}olmogorov's third-order structure-function
  law, the local isotropy hypothesis and the pressure-velocity correlation.
  {\em J. Fluid Mech.\/} {\bf 326}, 343--356.

\bibitem[{L'vov}(1991)]{lvov91}
{\sc {L'vov}, V.~S.} 1991 {Spectra of velocity and temperature fluctuations
  with constant entropy flux of fully developed free-convective turbulence}.
  {\em Phys. Rev. Lett.\/} {\bf 67}, 687--690.

\bibitem[{Mestayer} {\em et~al.\/}(1976){Mestayer}, {Gibson}, {Coantic} \&
  {Patel}]{mestayer76}
{\sc {Mestayer}, P., {Gibson}, C.~H., {Coantic}, M. \& {Patel}, A.} 1976 Local
  anisotropy in heated and cooled turbulent boundary layers. {\em Phys.
  Fluids\/} {\bf 19}, 1279--1287.

\bibitem[Monin \& Yaglom(1975)]{monin75}
{\sc Monin, A.~S. \& Yaglom, A.~M.} 1975 {\em Statistical fluid mechanics\/}, ,
  vol.~2. MIT Press.

\bibitem[{Obukhov}(1959)]{obukhov59}
{\sc {Obukhov}, A.~M.} 1959 {\em Dokl. Akad. Nauk. SSR\/} {\bf 125}, 1246.

\bibitem[{Parodi} {\em et~al.\/}(2004){Parodi}, {von Hardenberg}, {Passoni},
  {Provenzale} \& {Spiegel}]{parodi04}
{\sc {Parodi}, A., {von Hardenberg}, J., {Passoni}, G., {Provenzale}, A. \&
  {Spiegel}, E.~A.} 2004 {Clustering of Plumes in Turbulent Convection}. {\em
  Phys. Rev. Lett.\/} {\bf 92}~(19), 194503.

\bibitem[{Procaccia} \& {Zeitak}(1989)]{procaccia89}
{\sc {Procaccia}, I. \& {Zeitak}, R.} 1989 {Scaling exponents in nonisotropic
  convective turbulence}. {\em Phys. Rev. Lett.\/} {\bf 62}, 2128.

\bibitem[van Reeuwijk {\em et~al.\/}(2005)van Reeuwijk, Jonker \&
  Hanjalic]{reeuwijk05}
{\sc van Reeuwijk, M., Jonker, H.~J.~J. \& Hanjalic, K.} 2005 Identification of
  the wind in {R}ayleigh-{B}\'enard convection. {\em Phys. Fluids\/} {\bf 17},
  051704.

\bibitem[{Rincon} {\em et~al.\/}(2005){Rincon}, {Ligni{\` e}res} \&
  {Rieutord}]{rincon05}
{\sc {Rincon}, F., {Ligni{\` e}res}, F. \& {Rieutord}, M.} 2005 {Mesoscale
  flows in large aspect ratio simulations of turbulent compressible
  convection}. {\em Astron. \& Astrophys.\/} {\bf 430}, L57--L60.

\bibitem[{Toomre} {\em et~al.\/}(1990){Toomre}, {Brummell}, {Cattaneo} \&
  {Hurlburt}]{toomre90}
{\sc {Toomre}, J., {Brummell}, N., {Cattaneo}, F. \& {Hurlburt}, N.~E.} 1990
  {Three-dimensional compressible convection at low Prandtl numbers}. {\em
  Comp. Phys. Comm.\/} {\bf 59}, 105--117.

\bibitem[{Verzicco} \& {Camussi}(2003)]{verzicco03}
{\sc {Verzicco}, R. \& {Camussi}, R.} 2003 Numerical experiments on strongly
  turbulent thermal convection in a slender cylindrical cell. {\em J. Fluid
  Mech.\/} {\bf 477}, 19--49.

\bibitem[{Warhaft}(2000)]{warhaft2000}
{\sc {Warhaft}, Z.} 2000 {Passive scalars in turbulent flows}. {\em Ann. Rev.
  Fluid Mech.\/} {\bf 32}, 203--240.

\bibitem[{Yakhot}(1992)]{yakhot92}
{\sc {Yakhot}, V.} 1992 {4/5 {K}olmogorov law for statistically stationary
  turbulence: application to high-{R}ayleigh-number {B}{\' e}nard convection}.
  {\em Phys. Rev. Lett.\/} {\bf 69}, 769.

\end{thebibliography}

\end{document}